\documentclass[twocolumn,pra,nobalancelastpage,aps,raggedbottom]{revtex4-2}
\usepackage{times}

\usepackage{amsmath,amsfonts,amssymb,graphicx,hyperref,epstopdf,xcolor,tikz,scalerel,natbib,array}
\usepackage{mathptmx,textcomp,braket}
\usepackage[T1]{fontenc}
\usepackage[utf8]{inputenc}
\usepackage{multirow}
\usepackage{array}

\definecolor{lime}{HTML}{A6CE39}
\DeclareRobustCommand{\orcidicon}
{
	\begin{tikzpicture} 
	\draw[lime, fill=lime] (0,0) circle [radius=0.15] node[white] {{\fontfamily{qag}\selectfont \tiny ID}};
	\draw[white, fill=white] (-0.0625,0.095) 	circle [radius=0.007];
	\end{tikzpicture}
	\hspace{-2.2mm}
}
\newcommand\orcidID[1]{\href{https://orcid.org/#1}{\orcidicon}}

\newcommand{\be}{\begin {equation}}
\newcommand{\ee}{\end {equation}}
\newcommand{\beqa}{\begin {eqnarray}}
\newcommand{\eeqa}{\end {eqnarray}}
\newcommand{\mb}{\mathbf}
\newcommand{\Sch}{Schr\"odinger }

\hypersetup{colorlinks,citecolor=blue,filecolor=black,linkcolor=blue,urlcolor=blue}

\begin{document}

\title{High-order harmonic generation by sub-cycle laser pulses and associated scaling laws}

\author{Amol R. Holkundkar\orcidID{0000-0003-3889-0910}}
\email[E-mail: ]{amol.holkundkar@pilani.bits-pilani.ac.in}

\author{Rambabu Rajpoot\orcidID{0000-0002-2196-6133}}
  
\author{Jayendra N. Bandyopadhyay\orcidID{0000-0002-0825-9370}}

\affiliation{Department of Physics, Birla Institute of Technology and Science - Pilani, Rajasthan,
333031, India}

\date{\today}

\begin{abstract}
We studied the high-harmonic generation by the interaction of sub-cycle laser pulses with the He atom. The sub-cycle pulses are modeled using the complex source vector beam model, which is an exact solution to Maxwell's equations and accurately models the sub-cycle field profiles. We observed that the harmonic cutoff could be extended with the variation of the sub-cycle pulse duration, mainly because of the inherent blueshift associated with shorter pulses. The scaling laws for the harmonic yield and harmonic cutoff energy with pulse duration and fundamental driver wavelengths are also deduced. Furthermore, a detailed wavelet analysis of harmonic generation by sub-cycle pulses is carried out. The appropriate filtering and superposition of the harmonics gave rise to a single attosecond pulse of duration $\sim 100$ as. 
 
\end{abstract}

\maketitle

\section{Introduction}

Over the last couple of decades, we are witnessing tremendous advancement in the field of laser-atom interaction and so the high-order harmonic generation (HHG), which in turn manifested in the development of the extreme ultraviolet (XUV) and soft x-ray radiation sources \cite{Liu2019SpecLett,Mairesse2003_Science}. The generation of the high-harmonics from the laser-atom interaction can be understood by a celebrated   \textit{three-step model} \cite{PhysRevLett.70.1599,Corkum1993_PRL},  wherein the HHG is described as a three-step process, ionization of the electron, free propagation of electron in laser field and finally the recombination of the electron with the parent ion. Additional kinetic energy is emitted in the form of the higher harmonics of the fundamental frequency of the interacting laser pulse. 

The higher harmonics of the fundamental laser pulses are very crucial for the generation of the electromagnetic pulses at an attosecond time scale \cite{Krausz2009_RMP,Hentschel2001_Nature,Corkum2007_NatPhy}, which promises a vast number of applications \cite{Chini2014_nat,Krausz2009_RMP,Heuser2016_PRA,Ayuso2018_JPhysB,Baykusheva2016_PRL,Reich2016_PRL}. As we witnessed the rapid development on both theoretical and experimental front,  the research in this field is also focused toward polarization control of the emitted harmonics \cite{Huo2021_PRA,Zhang2017_OptLett,Rajpoot2021_JPhysB}, extending the cutoff energy of the HHG and increasing the corresponding intensity of the emitted high-order harmonics \cite{Astiaso2016}. In order to extend the higher harmonic cutoffs, different pulse shaping techniques have been introduced in the past \cite{PhysRevA.97.053414, PhysRevLett.117.093003,PhysRevA.96.033407,PhysRevA.93.033404,Rajpoot_2020,Holkundkar_2020}. The high-order harmonic generation from long pulses is very well understood in terms of the multiple rescattering \cite{Garcia2016,Lixin2014,Zhang_2016,Rost2022}. However, after the advent of intense ultrafast laser pulses \cite{Takahashi-2013,Emma-2004,Ding-2015}, the generation of the sub-cycle laser pulses is feasible, and now the study of quantum dynamics in ultra-short laser pulses is a field of contemporary interest around the globe \cite{Rossi2020,Chu_2016,Liang2017}. 

The spectral properties of the laser pulse play a crucial role in the scaling of high-order harmonic generation yield \cite{Tate2007_scale, wang2014_scale, Emelina2019}. Some fine oscillatory features in the wavelength scaling of harmonic yield are understood on the basis of quantum interference of many paths \cite{Schiessl2007_scale,Ishikawa2009_scale}. Furthermore, the experimental study on intensity-dependent conversion efficiency in HHG \cite{Nefedova2018} also corroborates the effect of laser spectral content. These scaling are also understood analytically using time-dependent effective range theory \cite{Frolov2009_scale}. Typically for mid-infrared lasers $\lambda_0 = 0.8\mu m\ \text{to}\ 2 \mu m$ the harmonic yield follows $\sim \lambda_0^{\gamma}$ scaling where $\lambda_0$ is the fundamental wavelength of the laser and $\gamma$ is the scaling parameter which varies from $\sim -4$ to $\sim -6$. In order to achieve high conversion efficiency, the experiments have demonstrated that for lower wavelength lasers, the harmonic conversion efficiency scales as $\gamma \sim -3$ \cite{Marceau_2017}. The inherent blueshift associated with the sub-cycle pulses \cite{LinSubcycle-2006} can be utilized to explore the possibilities of having enhanced harmonic yield, which in turn would be useful to generate narrowband XUV radiation for photoelectron spectroscopy of various atomic and molecular specimen \cite{Marceau_2017}.

Previously the sub-cycle pulses are used to study the harmonic generation in Hydrogen atom by solving the TDSE in \textit{one}-dimension \cite{Zheng_2011} to study the pulse envelope effects on the harmonic cutoffs. In this work, we explore the high-order harmonic generation by the linearly polarized sub-cycle laser pulse interacting with the Helium atom, and associated scaling laws for the harmonic cutoff and harmonic yields are also deduced as a function of the pulse duration and the fundamental driver wavelength. We solved the time-dependent \Sch equation (TDSE) through the time-dependent generalized pseudo-spectral (TDGPS) method in spherical coordinates. Routinely laser pulses longer than a few cycles have been represented as the product of a carrier wave and an envelope; however, in the sub-cycle regime, this definition breaks down as it acquires a dc component that can not be associated with the electromagnetic field of propagating light. To remedy this, we rely on the sub-cycle pulsed beam (SCPB) model, which is the exact solution of Maxwell's equations \cite{LinSubcycle-2006}.

The paper is organized as follows. Details of the numerical methods are discussed in Sec. \ref{sec2}, followed by the HHG scaling laws and attosecond pulse generation in Sec. \ref{sec3A} and \ref{sec3B}, respectively. The concluding remarks and future directions are discussed in Sec. \ref{sec4}. 
 

\section{Numerical Methods}
\label{sec2} 

\begin{figure}[t]
\centering\includegraphics[totalheight=0.35\textwidth]{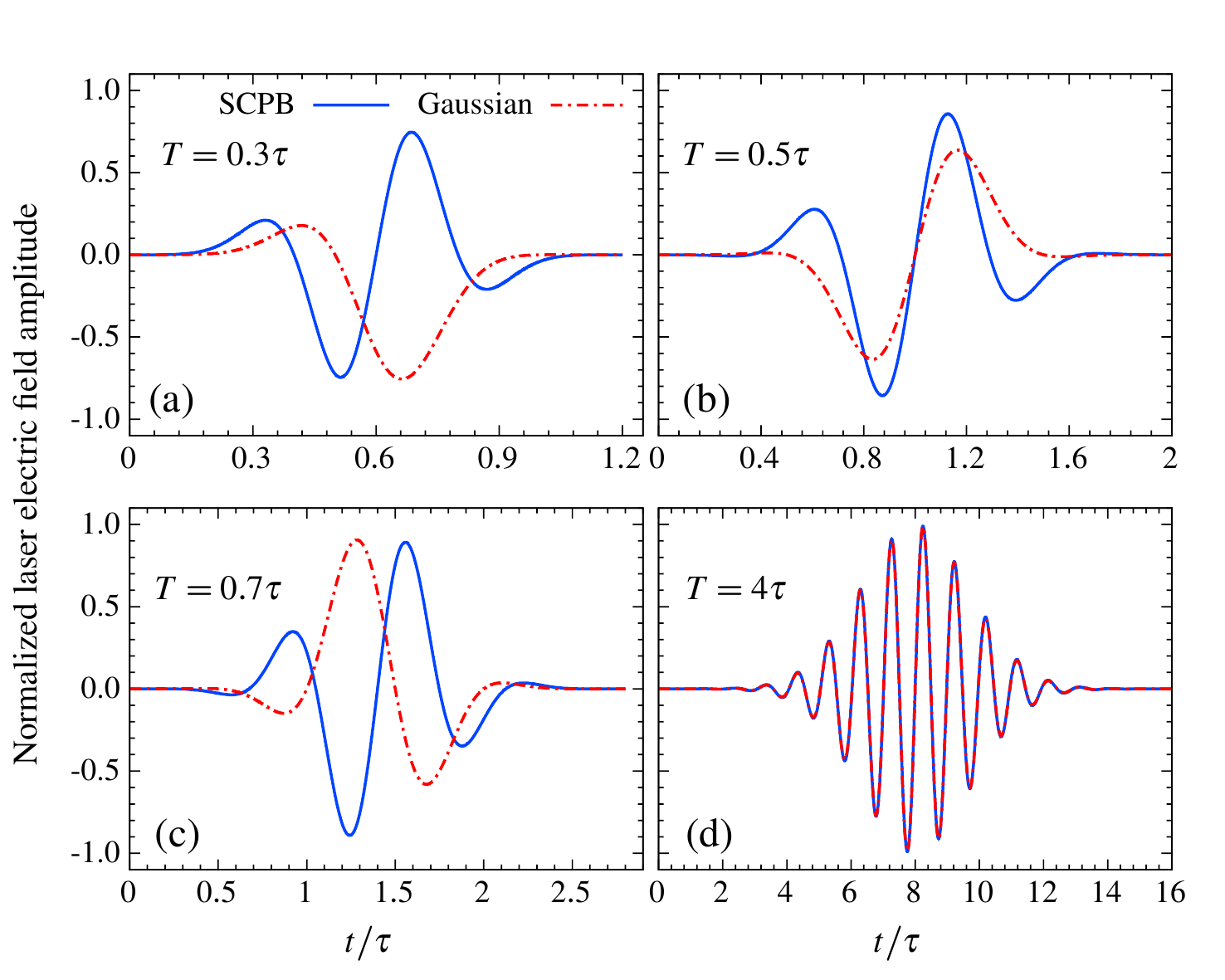}
\caption{Comparison of the normalized field profiles for SCPB \cite{LinSubcycle-2006} and Gaussian envelope [Eq. \ref{gaussLaser}] for FWHM pulse duration (a) $T = 0.3\tau$, (b) $T = 0.5\tau$, (c) $T = 0.7\tau$ and (d) $T = 4\tau$, where $\tau = \lambda/c$ is one laser cycle.  } 
\label{fig1}
\end{figure}

We study the interaction of the linearly polarized laser pulse with a He atom by numerically solving the TDSE under single-active-electron (SAE) approximation using TDGPS method \cite{TONG1997119}. The interaction of the linearly polarized laser ($m= 0$) with the spherically symmetric initial state of the He atom (1s state) will not alter the temporal evolution of the wavefunction in the azimuthal direction, and so effectively, it would be sufficient to solve the TDSE in radial and polar coordinates only. The TDSE in the length gauge is written as [atomic units are used throughout the manuscript] :
\be
	i \frac{\partial}{\partial t} \psi(\mb{r},t) = \big[- \frac{1}{2} \nabla^2 + V(r) + \mb{r}.\mb{E}(t) \big] \psi(\mb{r},t) ,
 \ee
where $\mb{E}(t)$ denotes the temporal profile of the linearly polarized laser pulse under dipole approximation. The linearly polarized Gaussian laser pulse is routinely expressed as:
\be \mb{E}(t) =  F_0 \sin(\omega_0 t) \exp\Big[-4 \ln 2 \Big(\frac{ t - 2 T}{T}\Big)^2 \Big]\  \hat{\mb{e}}_z \label{gaussLaser}\ee
wherein, $F_0 [\text{a.u.}] \sim 5.342 \times 10^{-9} \sqrt{I_0}$ is the electric field amplitude and $I_0$ is the laser intensity in W cm$^{-2}$, $\omega_0$ is the central frequency of the laser pulse and $T$ is the FWHM pulse duration with total simulation time considered to be $4T$. This plane wave model denoted by Eq. \ref{gaussLaser} is appropriate only for laser pulses longer than a few cycles; however, in the sub-cycle regime, this model seizes to be adequate, giving unrealistic field profiles. In order to circumvent this difficulty, we used the sub-cycle pulsed beam (SCPB)  \cite{LinSubcycle-2006} to express the sub-cycle field profiles. The SCPB is based on the complex-source model \cite{PhysRevE.67.016503,Heyman-89}, wherein an oscillating dipole is considered, which emits a spherical outgoing electromagnetic wave. A focused pulse is obtained by shifting the source dipole from origin to a complex position along the propagation direction. The fields so obtained are exact solutions of Maxwell's equations \cite{LinSubcycle-2006}. The SCPB is appropriate to model the laser fields under paraxial approximation and under very tight focusing (lower beam waist); it also accurately calculates the field along the propagation direction as well. Nevertheless, we restrict ourselves to dipole approximations wherein all spatial dependence in the field profiles is ignored, and only temporal dependence of the sub-cycle pulse is considered at the origin of the coordinate system (where field intensity maximizes). Moreover, the beam waist is considered to be large enough so that the longitudinal (along propagation direction) fields can also be ignored. We have compared the normalized field profiles for both SCPB and Gaussian envelope [Eq. \ref{gaussLaser}] in Fig. \ref{fig1} for different FWHM ($T$) laser pulse duration, and it can be observed that for sub-cycle pulses, the model given by Eq. \ref{gaussLaser} does not compute the field profile accurately, however for multi-cycle laser pulses the SCPB and plane wave model are in excellent agreement as expected. 

\begin{figure}[!b]
\centering\includegraphics[totalheight=0.8\columnwidth]{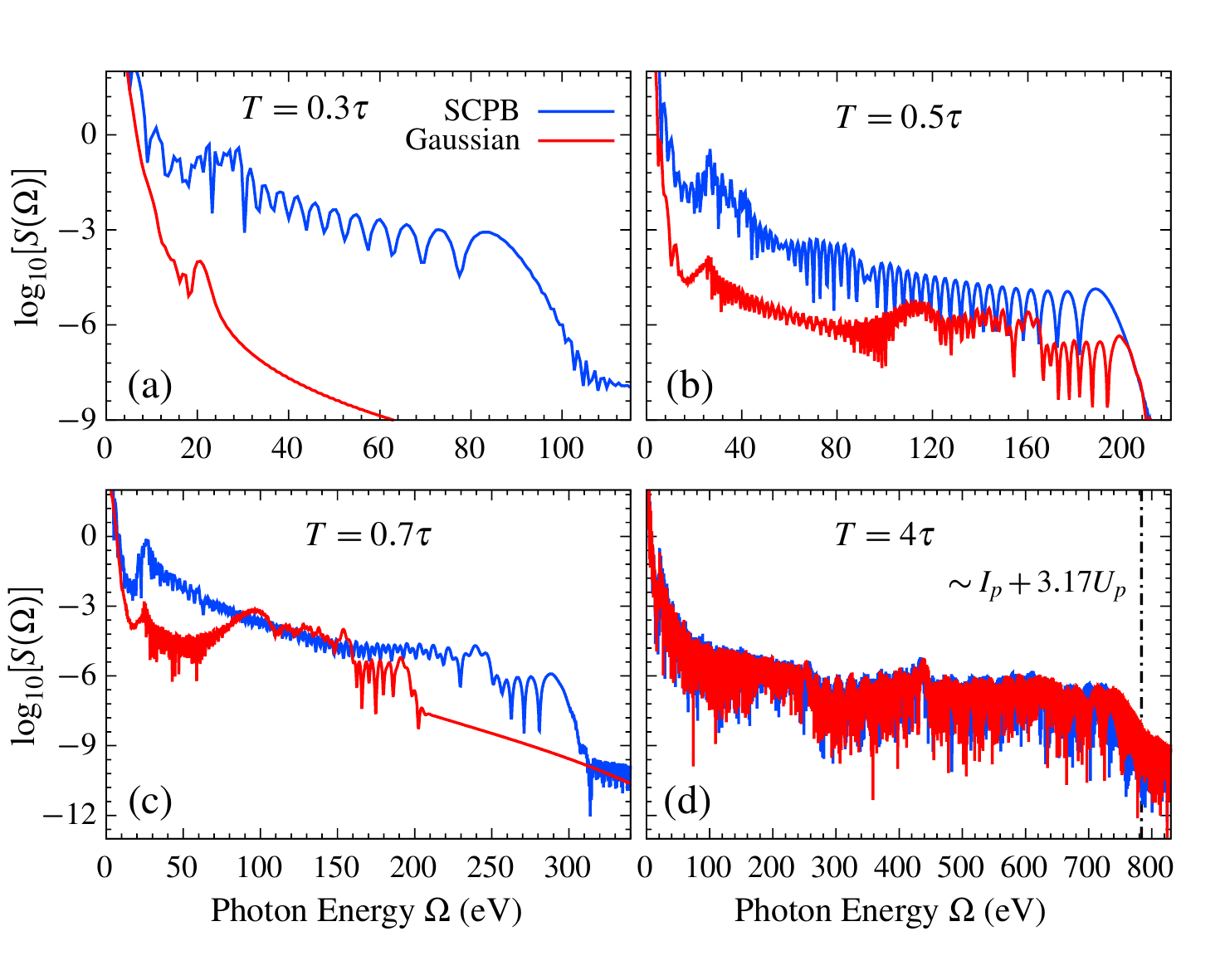}
\caption{Harmonic spectra with SCPB and plane wave Gaussian envelope is presented for pulse duration $0.3\tau$ (a), $0.5\tau$ (b), $0.7\tau$ (c) and $4\tau$ (d). The harmonic cutoff at $\sim 783$ eV as calculated by the three-step model is also illustrated in (d) as a dashed line. For all the cases, a 1600 nm laser is used with a peak intensity 10$^{15}$ W cm$^{-2}$.   } 
\label{fig2}
\end{figure}
 
The atomic Coulomb potential $V(r)$ for He atom under SAE is modeled by an empirical expression \cite{Tong_2005}, wherein the coefficients in the empirical expression of $V(r)$ are obtained by the self-interaction free density functional theory. The ground state (initial state) energy of the He is so obtained to be $\sim -0.9038$ a.u. after diagonalizing the Hamiltonian \cite{TONG1997119}. 
 
After solving the TDSE through TDGPS, the time-dependent dipole acceleration $\mb{a}(t)$ is evaluated following the Ehrenfest theorem as \cite{sandPRL_1999}:
\be
	\mb{a}(t) = - \Big\langle \psi(\mb{r},t) \Big| \frac{\partial V(r)}{\partial r} + \mb{E}(t) \Big| \psi(\mb{r},t) \Big\rangle.
 \ee
The harmonic spectra is then obtained by performing the Fourier transform of $\mb{a}(t)$ \cite{TONG1997119}, i.e.,
\be S(\Omega) \sim \Big|\frac{a(\Omega)}{\Omega^2} \Big|^2,\ee
where,
\be a(\Omega) = \frac{1}{\sqrt{2\pi}} \int \mb{a}(t) e^{-i\Omega t} dt\ee 
is the Fourier transform of the dipole acceleration. The integrated harmonic yield between the energies $E_1$ and $E_2$ are calculated as \cite{Ishikawa2009_scale,Schiessl2007_scale}, 
\be Y = \frac{1}{T} \int_{E_1}^{E_2} |a(\Omega)|^2 d\Omega \label{yield}\ee
where, $T$ is the pulse duration, $E_1$ and $E_2$ is the energy window in the plateau region. Furthermore, The field profile of the attosecond pulse [$\mathcal{E}_{asp}(t)$] can be constructed by filtering the desired frequency range using appropriate window function $w(\Omega)$ and is given as \cite{Peng-2020},
\be \mathcal{E}_{asp}(t) = \frac{1}{\sqrt{2\pi}} \int a(\Omega) w(\Omega) e^{i\Omega t} d\Omega.\ee 
The intensity of the attosecond pulse is then given by $I(t) \sim |\mathcal{E}_{asp}(t)|^2$. 
 
We considered the radial simulation domain of $r_{\text{max}} = 250$ a.u. and last $30$ a.u. is used as masking to absorb the outgoing wavefunction \cite{TONG1997119}. We mapped the radial domain $0\leq r \leq r_\text{max}$ to $-1 \leq r_p\leq 1$, such that $ r \equiv N ( 1 + r_p)/(1 - r_p + 2/r_\text{max})$, and $N$ is number of grid points. This mapped domain is divided in $N = 600$ non-uniform grid points, which are nothing but the roots of the Legendre polynomials, and hence there is no need to use any softening parameter in the Coulomb potential as required in Cartesian TDSE solvers to avoid singularity at $r = 0$. This is so because the singularity at $r = 0$ would correspond to $r_p = -1$, which will never be the root of the Legendre polynomial. Furthermore, maximum angular momentum is considered to be 50, and simulation time step $\sim 0.05$ a.u. are used. The TDGPS method enables the calculation of bound state wave functions and energies with very great precision. Our simulation utilizes widely used \textit{Armadillo} library for linear algebra purpose \cite{Sanderson2016}.

\begin{figure}[!t]
\centering\includegraphics[totalheight=0.8\columnwidth]{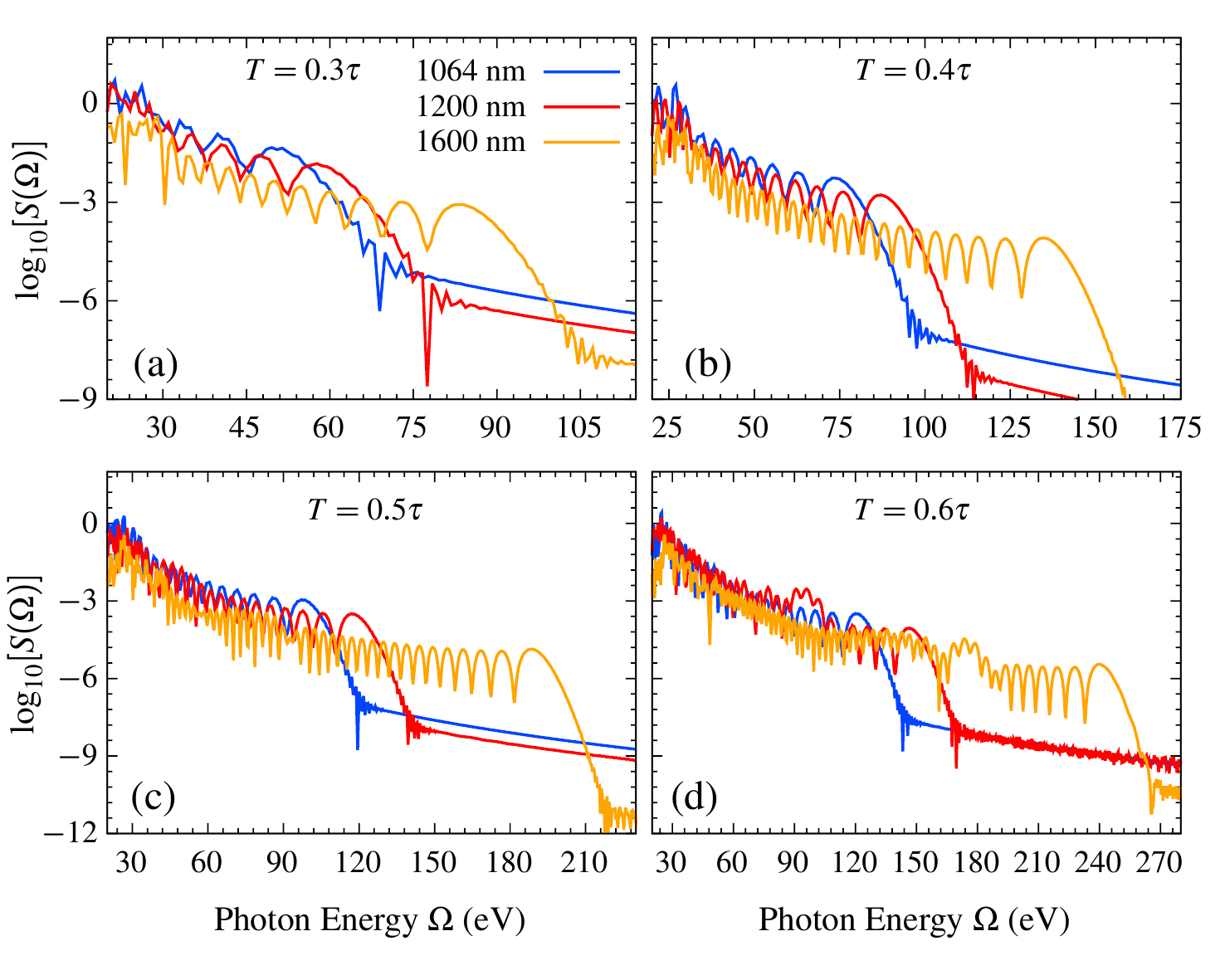}
\caption{Harmonic spectra using different wavelengths for various sub-cycle pulse duration is presented. The pulse duration of $0.3\tau$ (a), $0.4\tau$ (b), $0.5\tau$ (c) and $0.6\tau$ (d) are considered with peak intensity of the laser pulse is considered to be 10$^{15}$ W cm$^{-2}$ for all the cases. } 
\label{fig3}
\end{figure} 


In the following sections, we discuss the harmonic generation by sub-cycle pulses, and associate scaling laws are deduced for harmonic yield and cutoff energies in Sec. \ref{sec3A} followed by the wavelet analysis and the attosecond pulse generation in Sec. \ref{sec3B}

\section{Harmonic generation and scaling laws}
\label{sec3A}

We considered the interaction of 1600 nm laser with a peak intensity of 10$^{15}$ W cm$^{-2}$ with He atom, and the harmonic spectra were evaluated using both SCPB and plane wave Gaussian pulse model for different pulse duration, and the results are presented in Fig. \ref{fig2}. For pulse duration of $0.3\tau$ [Fig. \ref{fig2}(a)] we observe that the harmonic cutoff as calculated using the SCPB model is around $\sim 90$ eV, however for Gaussian pulse [Eq. \ref{gaussLaser}] the high-order harmonic generation seize to exist, which can be understood from the field profile of the Gaussian pulse as shown in Fig. \ref{fig1}(a). We observe from Fig. \ref{fig1}(a) that, for Gaussian pulse, the field profiles are not symmetric so that the ionized electron can recombine with the parent ion giving a higher harmonic. As we increase the pulse duration to say $0.5\tau$ and $0.7\tau$, the harmonic cutoff for SCPB is observed at $\sim 187$ eV and $\sim 288$ eV, respectively. For Gaussian pulse, the harmonic cutoff is observed around $\sim 193$ eV, with multi-plateau structure and different harmonic efficiency, which is attributed to the CEP of the laser envelope and wider spectral content of the laser, is it an ultra-short pulse. Moreover, as expected for pulse duration $4\tau$, the HHG spectra using the SCPB and Gaussian pulse model are identical, as the SCPB model converges to the Eq. \ref{gaussLaser} for longer pulse duration. The harmonic cutoff as predicted by the three-step model is $I_p + 3.17 U_p \sim 28.8\ \text{a.u.} \sim 783\ \text{eV}$ with $I_p = 0.9038$ a.u. and $U_P \sim F_0^2/4\omega_0^2 \sim 8.8\ \text{a.u.}$ are respectively the ionization potential and the ponderomotive energy. 

\begin{figure}[!t]
\centering\includegraphics[totalheight=0.8\columnwidth]{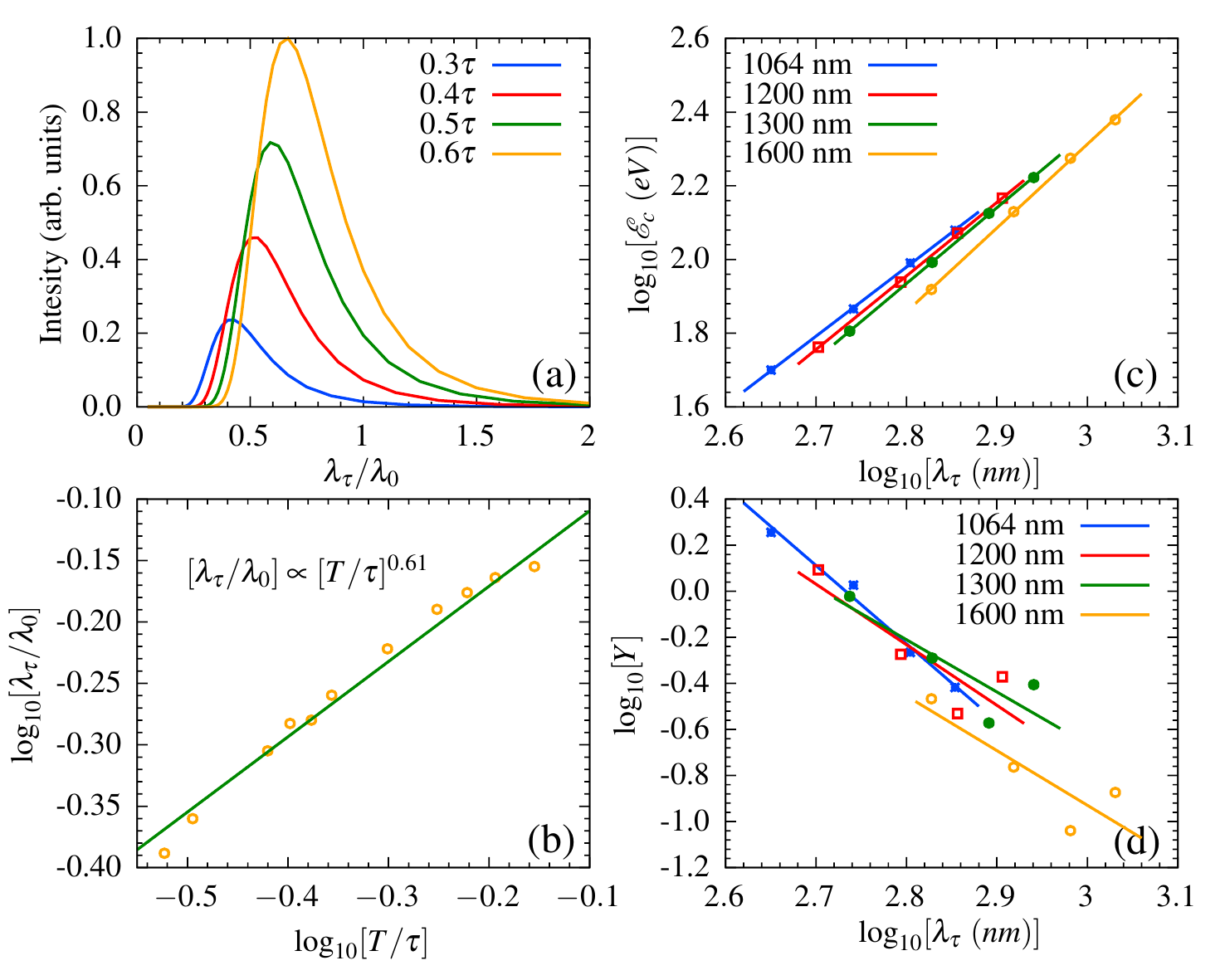}
\caption{The Fourier spectra of the sub-cycle pulses are presented (a), wherein the spectra are scaled with respect to the fundamental laser wavelength $\lambda_0$. The blueshift of the pulse [for a given fundamental laser wavelength] with pulse duration is illustrated in (b). The variation of harmonic cutoff energy $\mathcal{E}_c$ for different pulse duration (expressed as blueshifted wavelengths $\lambda_\tau$) is presented for different laser wavelengths $\lambda_0 = 1064, 1200, 1300$ and 1600 nm (c). Also the integrated harmonic yield for the energy range $0.25 \mathcal{E_\tau} - 0.75 \mathcal{E_\tau}$ (of $0.3\tau$) is also illustrated for different $\lambda_0$ (d). Please refer text for details on the scaling. In this figure, the dots represent the data obtained by simulation, and solid lines are the fitting. In (b), (c), and (d), points represent computed results, and the solid line represents scaling.} 
\label{fig4}
\end{figure}  

We also witness the harmonic cutoff extension with pulse duration for SCPB pulse [refer Fig. \ref{fig2}] \cite{Zheng_2011}. In order to further explore the effect of the pulse duration and laser wavelength, in Fig. \ref{fig3} we present the harmonic spectra for different pulse duration and wavelengths; for all the cases, the peak intensity of the laser pulse is considered to be 10$^{15}$ W cm$^{-2}$. For a particular pulse duration, the harmonic cutoff is observed to be enhanced by increasing the laser pulse wavelength. However, for fixed laser wavelength, the harmonic cutoff is enhanced by increasing the pulse duration as well. The variation in the pulse duration and the wavelength of the laser affects the Harmonic yield as well.

The enhancement in the harmonic cutoff with the pulse duration for a given laser wavelength can be understood by performing the Fourier analysis of the respective laser pulse. In Fig. \ref{fig4}(a) we present the FFT of the laser pulse for pulse duration $T = 0.3\tau$, $0.4\tau$, $0.5\tau$ and $0.6\tau$, and it can be observed that shorter pulses tend to be blueshifted with respect to the fundamental frequency (wavelength) of the laser $\omega_0$ ($\lambda_0$), and we denote the blueshifted wavelength by $\lambda_\tau$ for a given pulse duration. The variation of the normalized blueshifted wavelength ($\lambda_\tau/\lambda_0$) in terms of the normalized pulse duration ($T/\tau$) is presented in Fig. \ref{fig4}(b), and it is observed that $\lambda_\tau/\lambda_0$ follows the $\propto [T/\tau]^{0.61}$ scaling in the range we are concerned in this manuscript. Now we will represent all the scaling in terms of the blueshifted wavelength $\lambda_\tau$. The harmonic cutoff variation [$\mathcal{E}_c \propto \lambda_\tau^\alpha$] for different fundamental laser wavelengths ($\lambda_0$) are presented in Fig. \ref{fig4}(c) as a function of  $\lambda_\tau$  and the  scaling is obtained. It is observed that the cutoff energy scales as $\propto \lambda_\tau^\alpha$, with $\alpha \sim 1.875 $ (1064 nm), 1.99 (1200 nm), 2.06 (1300 nm), and 2.27 (1600 nm). This scaling of cutoff energy with increasing the pulse duration (reducing the blueshift) is mainly because of the increase in the ponderomotive energy, which scales as the square of the wavelength for a monochromatic electromagnetic wave (long pulse duration). We have also calculated the integrated harmonic yield for different fundamental laser wavelengths, and the results are presented in Fig. \ref{fig4}(d). The harmonic yield is calculated for the energy range $E_1 = 0.25 \mathcal{E}_{0.3\tau}$ and  $E_2 = 0.75 \mathcal{E}_{0.3\tau}$, wherein $\mathcal{E}_{0.3\tau}$ represents the cutoff energy for the $0.3\tau$ case for a particular laser wavelength $\lambda_0$, e.g. say for $\lambda_0 = 1600$ nm case the $0.3\tau$ pulse duration will be having the lowest cutoff at $\sim 83.2$ eV [Fig. \ref{fig3}(a)] and hence for 1600 nm case in Fig. \ref{fig4}(c) the harmonic yield [Eq. \ref{yield}] is calculated by integrating in the energy range $E_1 = 0.25 \times 83.2 \sim 21$ eV and $E_2 = 0.75 \times 83.2 \sim 62$ eV for all the pulse duration $0.4\tau$, $0.5\tau$ and $0.6\tau$, which we have presented in terms of the blueshifted wavelengths $\lambda_\tau$. The integrated harmonic yield is observed to follow the scaling $\propto \lambda_\tau^\beta$, where $\beta = -3.39$ (1064 nm), -2.63 (1200 nm), -2.27 (1300 nm), and -2.37 (1600 nm) is obtained.   
 
\begin{figure}[!b]
\centering\includegraphics[totalheight=0.55\columnwidth]{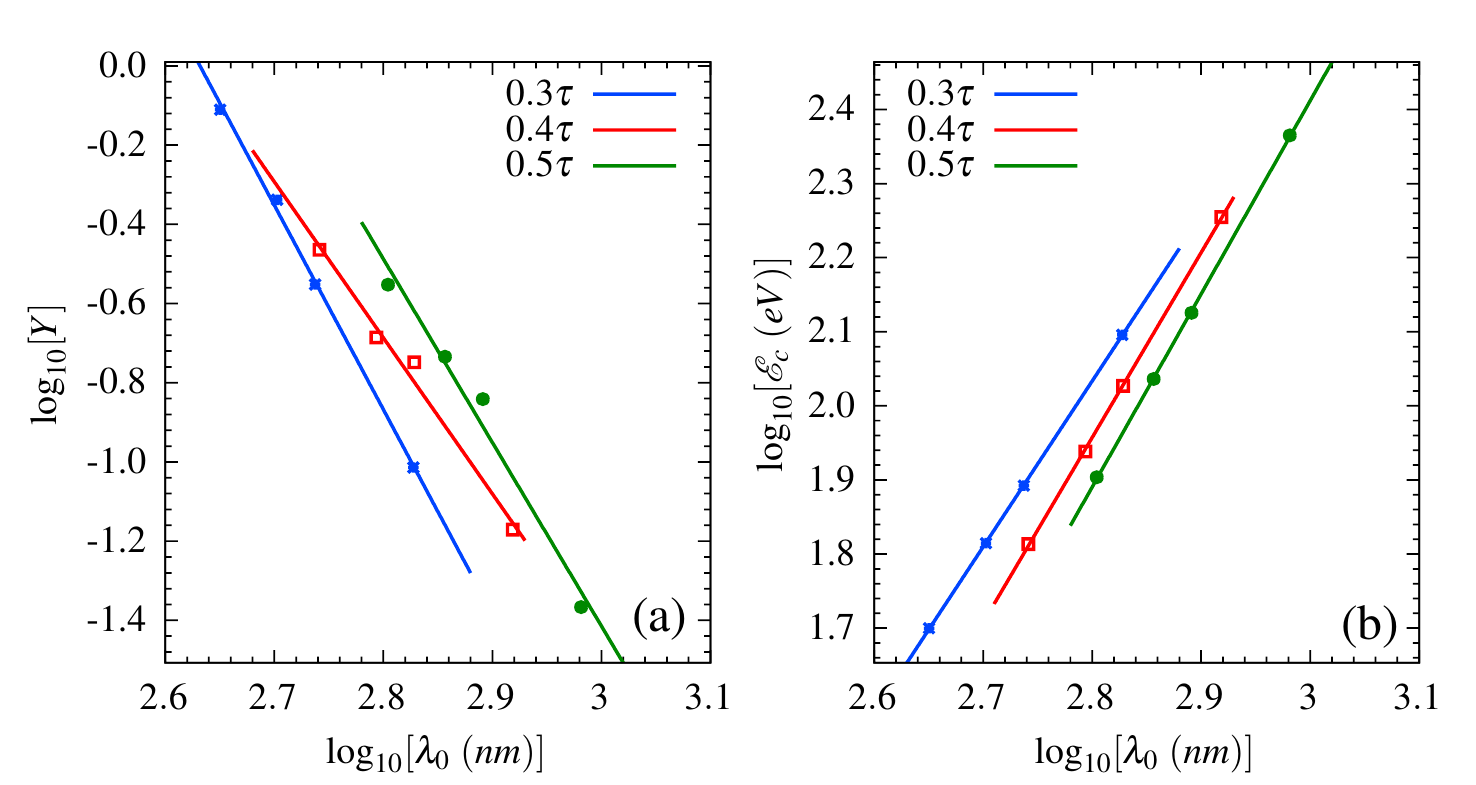}
\caption{Scaling of integrated harmonic yield from 30 - 50 eV (a) and cutoff energy (b) with fundamental laser wavelength using different pulse duration. Again the peak laser intensity is $10^{15}$ W cm$^{-2}$. Here points represent computed results, and a solid line represents scaling.} 
\label{fig5}
\end{figure}  
 
In Fig. \ref{fig5}(a) we present the integrated harmonic yield in the energy range 30 - 50 eV and harmonic cutoff energy as a function of the fundamental laser wavelength $\lambda_0$ for different pulse duration. The integrated harmonic yield is observed to follow the scaling $\propto \lambda_0^\gamma$ where, $\gamma = -5.16$ ($0.3\tau$), -3.94 ($0.4\tau$) and -4.63 ($0.5\tau$) is obtained. The scaling of harmonic yield with the laser wavelengths are studied extensively in the past  \cite{wang2014_scale,Frolov2009_scale, Schiessl2007_scale,Tate2007_scale}, wherein the harmonic yield scales as $\sim \lambda_0^\gamma$  with $-6 \leq \gamma \leq -4$ for the constant intensity of the laser pulse. The scaling presented in Fig. \ref{fig5}(a) is in line with previously reported numerical and analytical studies on the wavelength scaling of harmonic yield \cite{Frolov2015, Tate2007_scale}. These scaling would depend on the energy range in which the yield is calculated along with the range of the laser wavelengths. Furthermore, the harmonic cutoff energy scaling with fundamental laser wavelength is presented in Fig. \ref{fig5}(b) and the cutoff energy scaled as $\propto \lambda_0^\chi$ where, $\chi \sim 2.24$ ($0.3\tau$), 2.5 ($0.4\tau$) and 2.6 ($0.5\tau$) is obtained. This slight shift in the scaling of harmonic cutoff with the pulse duration can be understood from the pondermotive energy associated with different pulse duration. For a shorter pulse duration, the blueshifted frequency would cause the lower pondermotive energy resulting in weaker scaling and vice-versa.  
 
\begin{figure}[!b]
\centering\includegraphics[totalheight=0.8\columnwidth]{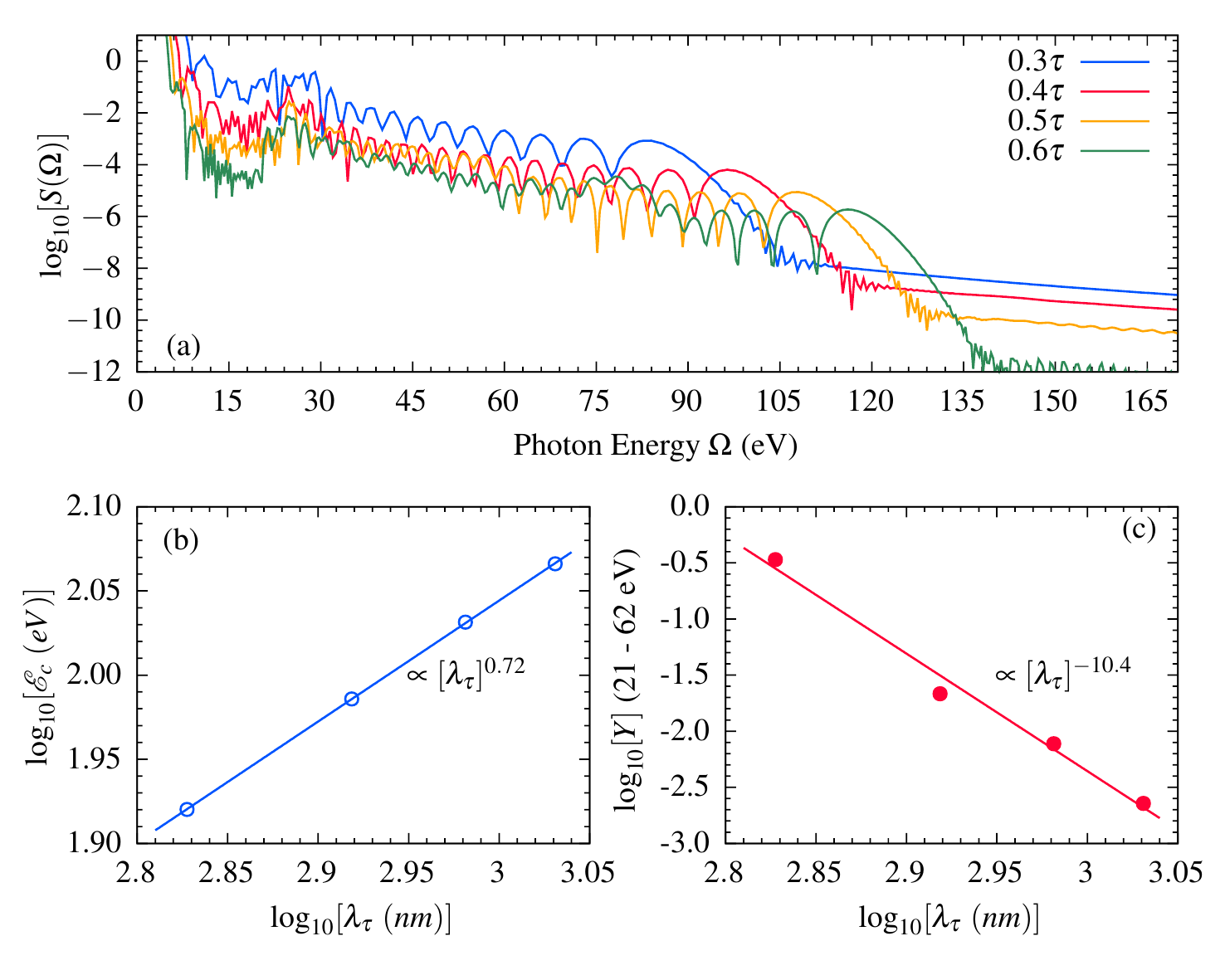}
\caption{Harmonic spectra is presented for different pulse duration for 1600 nm laser pulse (a), the peak intensity for $0.3\tau$ duration pulse is considered to be $10^{15}$ W cm$^{-2}$, and for other pulse duration, the intensity is lowered to conserve the energy of the pulse [$\propto \int |E(t)|^2 dt$]. Scaling for the cutoff energy (b) and integrated harmonic yield [21 - 62 eV] (c) are presented with respect to the blueshifted wavelength $\lambda_\tau$. The respective scaling parameters are also illustrated in (b) and (c). In (b) and (c), points represent computed results and solid line represents scaling. } 
\label{fig6}
\end{figure}

So far, while obtaining the harmonic cutoff and yield for different laser pulse duration [Fig. \ref{fig4}], we fixed the intensity of the laser pulse to $10^{15}$ W cm$^{-2}$, which translates the higher pulse energy for longer pulses and vice versa. Now, we obtain the harmonic spectra for these different pulse duration by keeping the pulse energy constant, and the results are presented in Fig. \ref{fig6}. Let us consider a case of 1600 nm laser pulse, with peak intensity $10^{15}$ W cm$^{-2}$ for $0.3\tau$ pulse duration; however, for $0.4\tau$, $0.5\tau$ and $0.6\tau$ the peak intensity is lowered such that the energy content of the pulse is constant. The energy-conserving pulse can be constructed by lowering the field amplitude such that $\sim \int |E(t)|^2 dt$ is constant for the different pulse duration pulses here $E(t)$ is the time-dependent field amplitude. We observed that the harmonic cutoff increases as $\propto [\lambda_\tau]^{0.72}$ where $\lambda_\tau$ is the blueshifted wavelength for a given pulse duration. However, the integrated harmonic yield between 21 - 62 eV is found to follow $\propto [\lambda_\tau]^{-10.4}$ scaling. If we compare this strong scaling with the one where the laser peak intensity is kept constant to 10$^{15}$ W cm$^{-2}$ for all the pulse duration, we obtained that the harmonic yield in the energy range 21 - 62 eV scaled as  $\propto [\lambda_\tau]^{-2.37}$ for 1600 nm laser pulse. These stronger scaling with the $\lambda_\tau$ for constant pulse energy case might be understood as the interplay between the intensity or field amplitude-dependent pondermotive energy and the wavelength-dependent harmonic yield. A further detailed study on how the intensity of sub-cycle pulses affects the scaling of the harmonic yield is presently beyond the purview of this manuscript. 


\begin{figure*}[t]
	\centering\includegraphics[totalheight=0.9\columnwidth]{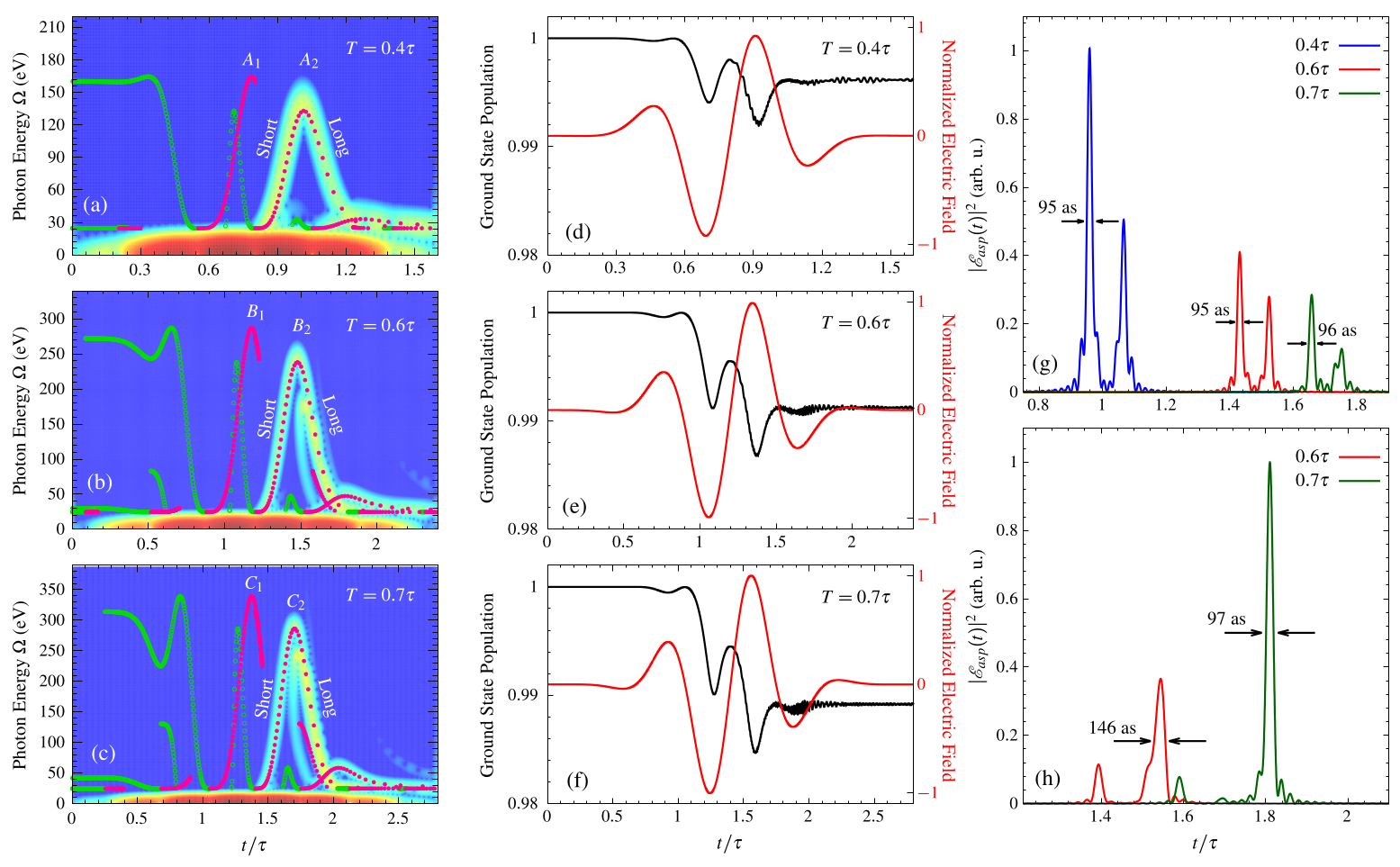}
	\caption{Left column: Time-frequency response of the dipole acceleration for different pulse duration: (a) $T = 0.4\tau$, (b) $T = 0.6\tau$, and (c) $T = 0.7\tau$. Classical recollision energies of the electron trajectories emitted for similar laser pulses are also shown for ionization (green open circles) and recombination (solid magenta circles) times, respectively. The classical recollision energies are up-shifted by $I_p = 24.59$ eV. Middle column: Ground state population (black curve) and the normalized laser electric field amplitude (red curve) for the SCPB duration: (d) $T = 0.4\tau$, (e) $T = 0.6\tau$, and (f) $T = 0.7\tau$. Right column: Temporal profile of the intensity $I(t) \sim |\mathcal{E}_{asp}(t)|^2$ of the ASPs constructed by superposing harmonics of: (g) energy range $40$ eV before cutoff energy $\mathcal{E}_c$ for different pulse duration, and (h) energy window $155-194$ eV for $T = 0.6\tau$, and $0.7\tau$ cases. For all the cases, sub-cycle pulses of wavelength $1600$ nm and peak intensity $10^{15}$ W cm$^{-2}$ are considered.} 
	\label{fig7}
\end{figure*}

\section{Attosecond pulse generation}
\label{sec3B}

Next, we present the Gabor transform for some representative cases to gain insight into the dynamics which leads to the structure in the harmonic spectrum. The results for pulse duration $0.4\tau$, $0.6\tau$, and $0.7\tau$ are presented in Fig. \ref{fig7} (a)-(c), respectively. Supplementarily, classical recollision energies of the electron trajectories emitted for similar laser pulses are also shown for ionization (green open circles) and recombination (solid magenta circles) times, respectively. In classical trajectories, the recombination energy up to the second return of the electron at the nucleus is considered. Also, the classical returning energies are up-shifted by the ionization potential of the He atom, i.e., $I_p = 24.59$ eV. In Fig. \ref{fig7} (d)-(f), we have shown the temporal variation in the ground state population (black curve) and the normalized laser electric field amplitude (red curve, right axis) for the SCPB duration $0.4\tau$, $0.6\tau$ and $0.7\tau$, respectively. For all the cases, sub-cycle laser pulses of wavelength $1600$ nm and peak field intensity $10^{15}$ W cm$^{-2}$ are considered. From Fig. \ref{fig7}(a) (the SCPB duration $T=0.4\tau$), we can see that in the time-frequency response, the photon emission is taking place between $0.8$ to $1.3$ laser cycle ($\tau$), which contributes to the HHG. The peak of this harmonic emission is located around $\sim 135$ eV, representing the harmonic cutoff which can also be seen in Fig. \ref{fig3}(b) (yellow curve). Moreover, there are two emission events taking place for each harmonic energy before the cutoff, corresponding to two different quantum paths, also referred to as the \textit{short} and \textit{long} trajectories. The interference of these two quantum paths is strong and leads to the modulations in HHG spectra. 

Furthermore, the HHG structure can be discussed using the classical trajectory analysis based on the semi-classical three-step model. In Fig. \ref{fig7}(a), there are two main peaks contributing to the HHG, marked as $A_1$ and $A_2$, respectively. The peak $A_2$ is originated due to the electron ionization between $0.65\tau$ and $0.8\tau$. The harmonic photon energy of peak $A_2$ is around $\sim 135$ eV with a return time between $0.8\tau$ to $1.3\tau$, which corresponds to the harmonic emission visible in the time-frequency response of the HHG. The peak $A_1$ is originated due to the recombination of electrons ionized before $0.55\tau$ and returned with the energy $\sim 165$ eV, which is missing in the time-frequency response. The plausible reason for this non-appearance can be given with the help of Fig. \ref{fig7}(d), wherein the temporal variation in ground state population and laser electric field amplitude are shown. From Fig. \ref{fig7}(d), It can be observed that the strength of the laser field (red curve) in the time window $0-0.55\tau$ is negligible, due to which the probability of ionization is insignificant before $0.55\tau$. Consequently, the harmonic efficiency is very low, and the subsequent emission of harmonics can be ignored, which is in agreement with the time-frequency response shown in Fig. \ref{fig7}(a). Figure \ref{fig7}(b) corresponding to $T = 0.6\tau$ case shows that the harmonic emission contributing to the HHG is taking place between $1.2\tau$ to $1.8\tau$. The cutoff energy of the HHG is located at $\sim 240$ eV. Similar to the $T=0.4\tau$ case, two emission events can be seen for each harmonic energy before the cutoff. The quantum path corresponding to a longer trajectory is less intense in the energy range $185$ eV to $240$ eV, which leads to a two plateau structure in the HHG spectrum as observed in Fig. \ref{fig3}(d) (yellow curve). The corresponding classical trajectory analysis for $T = 0.6\tau$ case shows two main peaks contributing to HHG, marked as $B_1$ and $B_2$, respectively. The peak $B_2$ is situated between $1.2\tau$ to $1.8\tau$ with maximum energy $\sim 240$ eV and corroborates with the harmonic emission peak present in the time-frequency response. However, there is no harmonic emission visible in the time-frequency analysis corresponding to the peak $B_1$, and the reason behind this disappearance can be given similar to the $T=0.4\tau$ case. The peak $B_1$ is the result of ionization occurring before time $0.85\tau$, wherein the laser field is quite weak, as can be seen in Fig. \ref{fig7}(e). As a result, the ground state population remains unchanged. Consequently, the HHG efficiency is so small that the associated harmonic emission can be neglected. Finally, we discuss the time-frequency response for the $T=0.7\tau$ case shown Fig. \ref{fig7}(c). The two different quantum paths are meeting at time $1.7\tau$ with cutoff energy $\sim 288$ eV. The longer path has less intensity in the energy range $250$ eV to $288$ eV, which is responsible for the two plateau structures in the HHG spectrum presented in Fig. \ref{fig2}(c) (blue curve). The corresponding classical trajectory analysis supports the Gabor analysis result, wherein the peak $C_2$ occurring between $1.4\tau$ to $2.1\tau$ with maximum energy $\sim 288$ eV corresponds to the harmonic emission. The reasoning behind the absence of the classical recollision energy peak $C_1$ in the Gabor transform is the same as for the $0.4\tau$ and $0.6\tau$ cases. We have also constructed the attosecond pulses using the high harmonics obtained for the cases $T=0.4\tau$, $0.6\tau$, and $0.7\tau$. 

The ASP intensity profiles $I \sim |\mathcal{E}_{asp}(t)|^2$ are also shown in Fig. \ref{fig7}(g) and (h) for these respective cases of $T=0.4\tau$, $0.6\tau$, and $0.7\tau$. In Fig. \ref{fig7}(g), the ASPs are constructed by superposing the harmonics of energy range $40$ eV before the cutoff energy $\mathcal{E}_c$ of different pulse duration cases. This implies that for $T=0.4\tau$, $0.6\tau$, and $0.7\tau$ cases, harmonics of energy windows $95-135$ eV, $200-240$ eV, and $248-288$ eV are selected, respectively. In all three cases, we observe two attosecond pulses corresponding to the two quantum emission paths. The ASP corresponding to the short quantum path has higher intensity compared to the long quantum path ASP in each case. This variation in intensity can be explained using the time-frequency responses shown in Fig. \ref{fig7}(a)-(c). In the time-frequency response, the short quantum path has higher intensity compared to the long path. This is reflected in the intensity of the attosecond pulses. The duration of the prominent ASP in each case is around 95 as, and the ASP intensity decreases with increasing pulse duration. The variation in attosecond pulse intensity corroborates with the scaling we have presented for the harmonic yield, in which the HHG yield decreases with increasing laser pulse duration. Moreover, a single attosecond pulse is also generated by properly choosing the energy range of filtered harmonics. In Fig. \ref{fig7}(h), single ASPs are shown for $T = 0.6\tau$, and $0.7\tau$ cases by selecting the harmonics of energy window $155-194$ eV. The obtained ASPs have duration $146$, and $97$ as for $T = 0.6\tau$, and $0.7\tau$ cases, respectively.

\section{Summary and Conclusions}
\label{sec4}

In summary, we have studied the high-order harmonic generation by sub-cycle laser pulse interacting with the He atom. For the sub-cycle pulse, we relied on the SCPB model, which is the exact solution of Maxwell's equations and which also avoids any ambiguities arising from the standard carrier envelope-based pulse models [Eq. \ref{gaussLaser}], however, for laser pulse longer than a few cycles, the SCPB model converges to  Eq. \ref{gaussLaser}. In this work, the harmonic generation is studied in the pulse duration range $0.3\tau \leq T \leq 0.7\tau$. It has been observed that the harmonic cutoff increases as the sub-cycle pulse duration increases; this is mainly because of the blueshift in the central frequency of the pulse with a reduction in pulse duration. Furthermore, we have also studied the effect of fundamental laser wavelength on the harmonic yield, and in the energy range 30 - 50 eV, the harmonic yield varies as $Y \propto \lambda_0^{-\alpha}$ with $4 \lesssim \alpha \lesssim 5.2$ for different laser pulse duration [Fig. \ref{fig5}(a)], these scaling with $\lambda_0$ are in line with the previously reported wavelength scaling. However, for a given $\lambda_0$ the variation of the harmonic yield with the pulse duration showed a weaker scaling $Y \propto \lambda_\tau^{-\alpha}$ such that $2.4 \lesssim \alpha \lesssim  3.4$ and $\lambda_\tau$ is the blueshifted wavelength for a particular pulse duration pulse, which suggests that for a given laser pulse one can have very robust scaling of harmonic yield with the variation of sub-cycle pulse duration, which in turn can be a source of narrowband XUV radiations. We also carried out the harmonic generation studies by keeping the laser pulse energy constant (field amplitude is reduced appropriately such that $\sim \int |E(t)|^2 dt$ is constant), and the harmonic yield (cutoff) showed stronger (weaker) scaling with the pulse duration as compare the case when the peak intensity of the laser is fixed for all duration. 

We also carried out a detailed wavelet analysis of the harmonic generation by sub-cycle pulses, and some detailed aspects of the harmonic spectra are explored using classical trajectories. The attosecond pulses are also synthesized by choosing suitable energy ranges in the harmonic plateau and generation of a single ASP of duration $\sim 100$ as is reported. The detailed study of the HHG with sub-cycle pulse intensities along with the macroscopic propagation effects \cite{Jin2011} we reserve for the future.

\section*{Acknowledgments} Authors would like to acknowledge the DST-SERB, Government of India, for funding the project CRG/2020/001020. 
  
\bibliographystyle{apsrev4-2}

\begin{thebibliography}{51}%
\makeatletter
\providecommand \@ifxundefined [1]{%
 \@ifx{#1\undefined}
}%
\providecommand \@ifnum [1]{%
 \ifnum #1\expandafter \@firstoftwo
 \else \expandafter \@secondoftwo
 \fi
}%
\providecommand \@ifx [1]{%
 \ifx #1\expandafter \@firstoftwo
 \else \expandafter \@secondoftwo
 \fi
}%
\providecommand \natexlab [1]{#1}%
\providecommand \enquote  [1]{``#1''}%
\providecommand \bibnamefont  [1]{#1}%
\providecommand \bibfnamefont [1]{#1}%
\providecommand \citenamefont [1]{#1}%
\providecommand \href@noop [0]{\@secondoftwo}%
\providecommand \href [0]{\begingroup \@sanitize@url \@href}%
\providecommand \@href[1]{\@@startlink{#1}\@@href}%
\providecommand \@@href[1]{\endgroup#1\@@endlink}%
\providecommand \@sanitize@url [0]{\catcode `\\12\catcode `\$12\catcode
  `\&12\catcode `\#12\catcode `\^12\catcode `\_12\catcode `\%12\relax}%
\providecommand \@@startlink[1]{}%
\providecommand \@@endlink[0]{}%
\providecommand \url  [0]{\begingroup\@sanitize@url \@url }%
\providecommand \@url [1]{\endgroup\@href {#1}{\urlprefix }}%
\providecommand \urlprefix  [0]{URL }%
\providecommand \Eprint [0]{\href }%
\providecommand \doibase [0]{https://doi.org/}%
\providecommand \selectlanguage [0]{\@gobble}%
\providecommand \bibinfo  [0]{\@secondoftwo}%
\providecommand \bibfield  [0]{\@secondoftwo}%
\providecommand \translation [1]{[#1]}%
\providecommand \BibitemOpen [0]{}%
\providecommand \bibitemStop [0]{}%
\providecommand \bibitemNoStop [0]{.\EOS\space}%
\providecommand \EOS [0]{\spacefactor3000\relax}%
\providecommand \BibitemShut  [1]{\csname bibitem#1\endcsname}%
\let\auto@bib@innerbib\@empty
\bibitem [{\citenamefont {Liu}\ and\ \citenamefont
  {Feng}(2019)}]{Liu2019SpecLett}%
  \BibitemOpen
  \bibfield  {author} {\bibinfo {author} {\bibfnamefont {H.}~\bibnamefont
  {Liu}}\ and\ \bibinfo {author} {\bibfnamefont {L.-Q.}\ \bibnamefont {Feng}},\
  }\href {https://doi.org/10.1080/00387010.2019.1585878} {\bibfield  {journal}
  {\bibinfo  {journal} {Spectroscopy Letters}\ }\textbf {\bibinfo {volume}
  {52}},\ \bibinfo {pages} {132} (\bibinfo {year} {2019})}\BibitemShut
  {NoStop}%
\bibitem [{\citenamefont {Mairesse}\ \emph {et~al.}(2003)\citenamefont
  {Mairesse}, \citenamefont {de~Bohan}, \citenamefont {Frasinski},
  \citenamefont {Merdji}, \citenamefont {Dinu}, \citenamefont {Monchicourt},
  \citenamefont {Breger}, \citenamefont {Kova{\v c}ev}, \citenamefont
  {Ta{\"\i}eb}, \citenamefont {Carr{\'e}}, \citenamefont {Muller},
  \citenamefont {Agostini},\ and\ \citenamefont
  {Sali{\`e}res}}]{Mairesse2003_Science}%
  \BibitemOpen
  \bibfield  {author} {\bibinfo {author} {\bibfnamefont {Y.}~\bibnamefont
  {Mairesse}}, \bibinfo {author} {\bibfnamefont {A.}~\bibnamefont {de~Bohan}},
  \bibinfo {author} {\bibfnamefont {L.~J.}\ \bibnamefont {Frasinski}}, \bibinfo
  {author} {\bibfnamefont {H.}~\bibnamefont {Merdji}}, \bibinfo {author}
  {\bibfnamefont {L.~C.}\ \bibnamefont {Dinu}}, \bibinfo {author}
  {\bibfnamefont {P.}~\bibnamefont {Monchicourt}}, \bibinfo {author}
  {\bibfnamefont {P.}~\bibnamefont {Breger}}, \bibinfo {author} {\bibfnamefont
  {M.}~\bibnamefont {Kova{\v c}ev}}, \bibinfo {author} {\bibfnamefont
  {R.}~\bibnamefont {Ta{\"\i}eb}}, \bibinfo {author} {\bibfnamefont
  {B.}~\bibnamefont {Carr{\'e}}}, \bibinfo {author} {\bibfnamefont {H.~G.}\
  \bibnamefont {Muller}}, \bibinfo {author} {\bibfnamefont {P.}~\bibnamefont
  {Agostini}},\ and\ \bibinfo {author} {\bibfnamefont {P.}~\bibnamefont
  {Sali{\`e}res}},\ }\href {https://doi.org/10.1126/science.1090277} {\bibfield
   {journal} {\bibinfo  {journal} {Science}\ }\textbf {\bibinfo {volume}
  {302}},\ \bibinfo {pages} {1540} (\bibinfo {year} {2003})}\BibitemShut
  {NoStop}%
\bibitem [{\citenamefont {Schafer}\ \emph {et~al.}(1993)\citenamefont
  {Schafer}, \citenamefont {Yang}, \citenamefont {DiMauro},\ and\ \citenamefont
  {Kulander}}]{PhysRevLett.70.1599}%
  \BibitemOpen
  \bibfield  {author} {\bibinfo {author} {\bibfnamefont {K.~J.}\ \bibnamefont
  {Schafer}}, \bibinfo {author} {\bibfnamefont {B.}~\bibnamefont {Yang}},
  \bibinfo {author} {\bibfnamefont {L.~F.}\ \bibnamefont {DiMauro}},\ and\
  \bibinfo {author} {\bibfnamefont {K.~C.}\ \bibnamefont {Kulander}},\ }\href
  {https://doi.org/10.1103/PhysRevLett.70.1599} {\bibfield  {journal} {\bibinfo
   {journal} {Phys. Rev. Lett.}\ }\textbf {\bibinfo {volume} {70}},\ \bibinfo
  {pages} {1599} (\bibinfo {year} {1993})}\BibitemShut {NoStop}%
\bibitem [{\citenamefont {Corkum}(1993)}]{Corkum1993_PRL}%
  \BibitemOpen
  \bibfield  {author} {\bibinfo {author} {\bibfnamefont {P.~B.}\ \bibnamefont
  {Corkum}},\ }\href {https://doi.org/10.1103/PhysRevLett.71.1994} {\bibfield
  {journal} {\bibinfo  {journal} {Phys. Rev. Lett.}\ }\textbf {\bibinfo
  {volume} {71}},\ \bibinfo {pages} {1994} (\bibinfo {year}
  {1993})}\BibitemShut {NoStop}%
\bibitem [{\citenamefont {Krausz}\ and\ \citenamefont
  {Ivanov}(2009)}]{Krausz2009_RMP}%
  \BibitemOpen
  \bibfield  {author} {\bibinfo {author} {\bibfnamefont {F.}~\bibnamefont
  {Krausz}}\ and\ \bibinfo {author} {\bibfnamefont {M.}~\bibnamefont
  {Ivanov}},\ }\href {https://doi.org/10.1103/RevModPhys.81.163} {\bibfield
  {journal} {\bibinfo  {journal} {Rev. Mod. Phys.}\ }\textbf {\bibinfo {volume}
  {81}},\ \bibinfo {pages} {163} (\bibinfo {year} {2009})}\BibitemShut
  {NoStop}%
\bibitem [{\citenamefont {Hentschel}\ \emph {et~al.}(2001)\citenamefont
  {Hentschel}, \citenamefont {Kienberger}, \citenamefont {Spielmann},
  \citenamefont {Reider}, \citenamefont {Milosevic}, \citenamefont {Brabec},
  \citenamefont {Corkum}, \citenamefont {Heinzmann}, \citenamefont {Drescher},\
  and\ \citenamefont {Krausz}}]{Hentschel2001_Nature}%
  \BibitemOpen
  \bibfield  {author} {\bibinfo {author} {\bibfnamefont {M.}~\bibnamefont
  {Hentschel}}, \bibinfo {author} {\bibfnamefont {R.}~\bibnamefont
  {Kienberger}}, \bibinfo {author} {\bibfnamefont {C.}~\bibnamefont
  {Spielmann}}, \bibinfo {author} {\bibfnamefont {G.~A.}\ \bibnamefont
  {Reider}}, \bibinfo {author} {\bibfnamefont {N.}~\bibnamefont {Milosevic}},
  \bibinfo {author} {\bibfnamefont {T.}~\bibnamefont {Brabec}}, \bibinfo
  {author} {\bibfnamefont {P.}~\bibnamefont {Corkum}}, \bibinfo {author}
  {\bibfnamefont {U.}~\bibnamefont {Heinzmann}}, \bibinfo {author}
  {\bibfnamefont {M.}~\bibnamefont {Drescher}},\ and\ \bibinfo {author}
  {\bibfnamefont {F.}~\bibnamefont {Krausz}},\ }\href
  {https://doi.org/10.1038/35107000} {\bibfield  {journal} {\bibinfo  {journal}
  {Nature}\ }\textbf {\bibinfo {volume} {414}},\ \bibinfo {pages} {509}
  (\bibinfo {year} {2001})}\BibitemShut {NoStop}%
\bibitem [{\citenamefont {Corkum}\ and\ \citenamefont
  {Krausz}(2007)}]{Corkum2007_NatPhy}%
  \BibitemOpen
  \bibfield  {author} {\bibinfo {author} {\bibfnamefont {P.~B.}\ \bibnamefont
  {Corkum}}\ and\ \bibinfo {author} {\bibfnamefont {F.}~\bibnamefont
  {Krausz}},\ }\href {https://doi.org/10.1038/nphys620} {\bibfield  {journal}
  {\bibinfo  {journal} {Nat. Phys.}\ }\textbf {\bibinfo {volume} {3}},\
  \bibinfo {pages} {381} (\bibinfo {year} {2007})}\BibitemShut {NoStop}%
\bibitem [{\citenamefont {Chini}\ \emph {et~al.}(2014)\citenamefont {Chini},
  \citenamefont {Zhao},\ and\ \citenamefont {Chang}}]{Chini2014_nat}%
  \BibitemOpen
  \bibfield  {author} {\bibinfo {author} {\bibfnamefont {M.}~\bibnamefont
  {Chini}}, \bibinfo {author} {\bibfnamefont {K.}~\bibnamefont {Zhao}},\ and\
  \bibinfo {author} {\bibfnamefont {Z.}~\bibnamefont {Chang}},\ }\href
  {http://dx.doi.org/10.1038/nphoton.2013.362} {\bibfield  {journal} {\bibinfo
  {journal} {Nat. Photon.}\ }\textbf {\bibinfo {volume} {8}},\ \bibinfo {pages}
  {178} (\bibinfo {year} {2014})}\BibitemShut {NoStop}%
\bibitem [{\citenamefont {Heuser}\ \emph {et~al.}(2016)\citenamefont {Heuser},
  \citenamefont {Jim\'enez~Gal\'an}, \citenamefont {Cirelli}, \citenamefont
  {Marante}, \citenamefont {Sabbar}, \citenamefont {Boge}, \citenamefont
  {Lucchini}, \citenamefont {Gallmann}, \citenamefont {Ivanov}, \citenamefont
  {Kheifets}, \citenamefont {Dahlstr\"om}, \citenamefont {Lindroth},
  \citenamefont {Argenti}, \citenamefont {Mart\'{\i}n},\ and\ \citenamefont
  {Keller}}]{Heuser2016_PRA}%
  \BibitemOpen
  \bibfield  {author} {\bibinfo {author} {\bibfnamefont {S.}~\bibnamefont
  {Heuser}}, \bibinfo {author} {\bibfnamefont {A.}~\bibnamefont
  {Jim\'enez~Gal\'an}}, \bibinfo {author} {\bibfnamefont {C.}~\bibnamefont
  {Cirelli}}, \bibinfo {author} {\bibfnamefont {C.}~\bibnamefont {Marante}},
  \bibinfo {author} {\bibfnamefont {M.}~\bibnamefont {Sabbar}}, \bibinfo
  {author} {\bibfnamefont {R.}~\bibnamefont {Boge}}, \bibinfo {author}
  {\bibfnamefont {M.}~\bibnamefont {Lucchini}}, \bibinfo {author}
  {\bibfnamefont {L.}~\bibnamefont {Gallmann}}, \bibinfo {author}
  {\bibfnamefont {I.}~\bibnamefont {Ivanov}}, \bibinfo {author} {\bibfnamefont
  {A.~S.}\ \bibnamefont {Kheifets}}, \bibinfo {author} {\bibfnamefont {J.~M.}\
  \bibnamefont {Dahlstr\"om}}, \bibinfo {author} {\bibfnamefont
  {E.}~\bibnamefont {Lindroth}}, \bibinfo {author} {\bibfnamefont
  {L.}~\bibnamefont {Argenti}}, \bibinfo {author} {\bibfnamefont
  {F.}~\bibnamefont {Mart\'{\i}n}},\ and\ \bibinfo {author} {\bibfnamefont
  {U.}~\bibnamefont {Keller}},\ }\href
  {https://doi.org/10.1103/PhysRevA.94.063409} {\bibfield  {journal} {\bibinfo
  {journal} {Phys. Rev. A}\ }\textbf {\bibinfo {volume} {94}},\ \bibinfo
  {pages} {063409} (\bibinfo {year} {2016})}\BibitemShut {NoStop}%
\bibitem [{\citenamefont {Ayuso}\ \emph {et~al.}(2018)\citenamefont {Ayuso},
  \citenamefont {Decleva}, \citenamefont {Patchkovskii},\ and\ \citenamefont
  {Smirnova}}]{Ayuso2018_JPhysB}%
  \BibitemOpen
  \bibfield  {author} {\bibinfo {author} {\bibfnamefont {D.}~\bibnamefont
  {Ayuso}}, \bibinfo {author} {\bibfnamefont {P.}~\bibnamefont {Decleva}},
  \bibinfo {author} {\bibfnamefont {S.}~\bibnamefont {Patchkovskii}},\ and\
  \bibinfo {author} {\bibfnamefont {O.}~\bibnamefont {Smirnova}},\ }\href
  {https://doi.org/10.1088/1361-6455/aaae5e} {\bibfield  {journal} {\bibinfo
  {journal} {J. Phys. B: At. Mol. Opt. Phys.}\ }\textbf {\bibinfo {volume}
  {51}},\ \bibinfo {pages} {06LT01} (\bibinfo {year} {2018})}\BibitemShut
  {NoStop}%
\bibitem [{\citenamefont {Baykusheva}\ \emph {et~al.}(2016)\citenamefont
  {Baykusheva}, \citenamefont {Ahsan}, \citenamefont {Lin},\ and\ \citenamefont
  {W\"orner}}]{Baykusheva2016_PRL}%
  \BibitemOpen
  \bibfield  {author} {\bibinfo {author} {\bibfnamefont {D.}~\bibnamefont
  {Baykusheva}}, \bibinfo {author} {\bibfnamefont {M.~S.}\ \bibnamefont
  {Ahsan}}, \bibinfo {author} {\bibfnamefont {N.}~\bibnamefont {Lin}},\ and\
  \bibinfo {author} {\bibfnamefont {H.~J.}\ \bibnamefont {W\"orner}},\ }\href
  {https://doi.org/10.1103/PhysRevLett.116.123001} {\bibfield  {journal}
  {\bibinfo  {journal} {Phys. Rev. Lett.}\ }\textbf {\bibinfo {volume} {116}},\
  \bibinfo {pages} {123001} (\bibinfo {year} {2016})}\BibitemShut {NoStop}%
\bibitem [{\citenamefont {Reich}\ and\ \citenamefont
  {Madsen}(2016)}]{Reich2016_PRL}%
  \BibitemOpen
  \bibfield  {author} {\bibinfo {author} {\bibfnamefont {D.~M.}\ \bibnamefont
  {Reich}}\ and\ \bibinfo {author} {\bibfnamefont {L.~B.}\ \bibnamefont
  {Madsen}},\ }\href {https://doi.org/10.1103/PhysRevLett.117.133902}
  {\bibfield  {journal} {\bibinfo  {journal} {Phys. Rev. Lett.}\ }\textbf
  {\bibinfo {volume} {117}},\ \bibinfo {pages} {133902} (\bibinfo {year}
  {2016})}\BibitemShut {NoStop}%
\bibitem [{\citenamefont {Huo}\ \emph {et~al.}(2021)\citenamefont {Huo},
  \citenamefont {Xing}, \citenamefont {Qi}, \citenamefont {Sun}, \citenamefont
  {Li}, \citenamefont {Zhang},\ and\ \citenamefont {Liu}}]{Huo2021_PRA}%
  \BibitemOpen
  \bibfield  {author} {\bibinfo {author} {\bibfnamefont {X.-X.}\ \bibnamefont
  {Huo}}, \bibinfo {author} {\bibfnamefont {Y.-H.}\ \bibnamefont {Xing}},
  \bibinfo {author} {\bibfnamefont {T.}~\bibnamefont {Qi}}, \bibinfo {author}
  {\bibfnamefont {Y.}~\bibnamefont {Sun}}, \bibinfo {author} {\bibfnamefont
  {B.}~\bibnamefont {Li}}, \bibinfo {author} {\bibfnamefont {J.}~\bibnamefont
  {Zhang}},\ and\ \bibinfo {author} {\bibfnamefont {X.-S.}\ \bibnamefont
  {Liu}},\ }\href {https://doi.org/10.1103/PhysRevA.103.053116} {\bibfield
  {journal} {\bibinfo  {journal} {Phys. Rev. A}\ }\textbf {\bibinfo {volume}
  {103}},\ \bibinfo {pages} {053116} (\bibinfo {year} {2021})}\BibitemShut
  {NoStop}%
\bibitem [{\citenamefont {Zhang}\ \emph {et~al.}(2017)\citenamefont {Zhang},
  \citenamefont {Zhu}, \citenamefont {Liu}, \citenamefont {Wang}, \citenamefont
  {Zhang}, \citenamefont {Lan},\ and\ \citenamefont {Lu}}]{Zhang2017_OptLett}%
  \BibitemOpen
  \bibfield  {author} {\bibinfo {author} {\bibfnamefont {X.}~\bibnamefont
  {Zhang}}, \bibinfo {author} {\bibfnamefont {X.}~\bibnamefont {Zhu}}, \bibinfo
  {author} {\bibfnamefont {X.}~\bibnamefont {Liu}}, \bibinfo {author}
  {\bibfnamefont {D.}~\bibnamefont {Wang}}, \bibinfo {author} {\bibfnamefont
  {Q.}~\bibnamefont {Zhang}}, \bibinfo {author} {\bibfnamefont
  {P.}~\bibnamefont {Lan}},\ and\ \bibinfo {author} {\bibfnamefont
  {P.}~\bibnamefont {Lu}},\ }\href {https://doi.org/10.1364/OL.42.001027}
  {\bibfield  {journal} {\bibinfo  {journal} {Opt. Lett.}\ }\textbf {\bibinfo
  {volume} {42}},\ \bibinfo {pages} {1027} (\bibinfo {year}
  {2017})}\BibitemShut {NoStop}%
\bibitem [{\citenamefont {Rajpoot}\ \emph {et~al.}(2021)\citenamefont
  {Rajpoot}, \citenamefont {Holkundkar},\ and\ \citenamefont
  {Bandyopadhyay}}]{Rajpoot2021_JPhysB}%
  \BibitemOpen
  \bibfield  {author} {\bibinfo {author} {\bibfnamefont {R.}~\bibnamefont
  {Rajpoot}}, \bibinfo {author} {\bibfnamefont {A.~R.}\ \bibnamefont
  {Holkundkar}},\ and\ \bibinfo {author} {\bibfnamefont {J.~N.}\ \bibnamefont
  {Bandyopadhyay}},\ }\href {https://doi.org/10.1088/1361-6455/ac3f97}
  {\bibfield  {journal} {\bibinfo  {journal} {J. Phys. B: At. Mol. Opt. Phys.}\
  }\textbf {\bibinfo {volume} {54}},\ \bibinfo {pages} {225401} (\bibinfo
  {year} {2021})}\BibitemShut {NoStop}%
\bibitem [{\citenamefont {Lara-Astiaso}\ \emph
  {et~al.}(2016{\natexlab{a}})\citenamefont {Lara-Astiaso}, \citenamefont
  {Silva}, \citenamefont {Gubaydullin}, \citenamefont {Rivi\`ere},
  \citenamefont {Meier},\ and\ \citenamefont {Mart\'{\i}n}}]{Astiaso2016}%
  \BibitemOpen
  \bibfield  {author} {\bibinfo {author} {\bibfnamefont {M.}~\bibnamefont
  {Lara-Astiaso}}, \bibinfo {author} {\bibfnamefont {R.~E.~F.}\ \bibnamefont
  {Silva}}, \bibinfo {author} {\bibfnamefont {A.}~\bibnamefont {Gubaydullin}},
  \bibinfo {author} {\bibfnamefont {P.}~\bibnamefont {Rivi\`ere}}, \bibinfo
  {author} {\bibfnamefont {C.}~\bibnamefont {Meier}},\ and\ \bibinfo {author}
  {\bibfnamefont {F.}~\bibnamefont {Mart\'{\i}n}},\ }\href
  {https://doi.org/10.1103/PhysRevLett.117.093003} {\bibfield  {journal}
  {\bibinfo  {journal} {Phys. Rev. Lett.}\ }\textbf {\bibinfo {volume} {117}},\
  \bibinfo {pages} {093003} (\bibinfo {year} {2016}{\natexlab{a}})}\BibitemShut
  {NoStop}%
\bibitem [{\citenamefont {Peng}\ \emph {et~al.}(2018)\citenamefont {Peng},
  \citenamefont {Frolov}, \citenamefont {Pi},\ and\ \citenamefont
  {Starace}}]{PhysRevA.97.053414}%
  \BibitemOpen
  \bibfield  {author} {\bibinfo {author} {\bibfnamefont {D.}~\bibnamefont
  {Peng}}, \bibinfo {author} {\bibfnamefont {M.~V.}\ \bibnamefont {Frolov}},
  \bibinfo {author} {\bibfnamefont {L.-W.}\ \bibnamefont {Pi}},\ and\ \bibinfo
  {author} {\bibfnamefont {A.~F.}\ \bibnamefont {Starace}},\ }\href
  {https://doi.org/10.1103/PhysRevA.97.053414} {\bibfield  {journal} {\bibinfo
  {journal} {Phys. Rev. A}\ }\textbf {\bibinfo {volume} {97}},\ \bibinfo
  {pages} {053414} (\bibinfo {year} {2018})}\BibitemShut {NoStop}%
\bibitem [{\citenamefont {Lara-Astiaso}\ \emph
  {et~al.}(2016{\natexlab{b}})\citenamefont {Lara-Astiaso}, \citenamefont
  {Silva}, \citenamefont {Gubaydullin}, \citenamefont {Rivi\`ere},
  \citenamefont {Meier},\ and\ \citenamefont
  {Mart\'{\i}n}}]{PhysRevLett.117.093003}%
  \BibitemOpen
  \bibfield  {author} {\bibinfo {author} {\bibfnamefont {M.}~\bibnamefont
  {Lara-Astiaso}}, \bibinfo {author} {\bibfnamefont {R.~E.~F.}\ \bibnamefont
  {Silva}}, \bibinfo {author} {\bibfnamefont {A.}~\bibnamefont {Gubaydullin}},
  \bibinfo {author} {\bibfnamefont {P.}~\bibnamefont {Rivi\`ere}}, \bibinfo
  {author} {\bibfnamefont {C.}~\bibnamefont {Meier}},\ and\ \bibinfo {author}
  {\bibfnamefont {F.}~\bibnamefont {Mart\'{\i}n}},\ }\href
  {https://doi.org/10.1103/PhysRevLett.117.093003} {\bibfield  {journal}
  {\bibinfo  {journal} {Phys. Rev. Lett.}\ }\textbf {\bibinfo {volume} {117}},\
  \bibinfo {pages} {093003} (\bibinfo {year} {2016}{\natexlab{b}})}\BibitemShut
  {NoStop}%
\bibitem [{\citenamefont {Wang}\ \emph {et~al.}(2017)\citenamefont {Wang},
  \citenamefont {Liu}, \citenamefont {He}, \citenamefont {Li}, \citenamefont
  {Wang}, \citenamefont {Zhu}, \citenamefont {Lan},\ and\ \citenamefont
  {Lu}}]{PhysRevA.96.033407}%
  \BibitemOpen
  \bibfield  {author} {\bibinfo {author} {\bibfnamefont {F.}~\bibnamefont
  {Wang}}, \bibinfo {author} {\bibfnamefont {W.}~\bibnamefont {Liu}}, \bibinfo
  {author} {\bibfnamefont {L.}~\bibnamefont {He}}, \bibinfo {author}
  {\bibfnamefont {L.}~\bibnamefont {Li}}, \bibinfo {author} {\bibfnamefont
  {B.}~\bibnamefont {Wang}}, \bibinfo {author} {\bibfnamefont {X.}~\bibnamefont
  {Zhu}}, \bibinfo {author} {\bibfnamefont {P.}~\bibnamefont {Lan}},\ and\
  \bibinfo {author} {\bibfnamefont {P.}~\bibnamefont {Lu}},\ }\href
  {https://doi.org/10.1103/PhysRevA.96.033407} {\bibfield  {journal} {\bibinfo
  {journal} {Phys. Rev. A}\ }\textbf {\bibinfo {volume} {96}},\ \bibinfo
  {pages} {033407} (\bibinfo {year} {2017})}\BibitemShut {NoStop}%
\bibitem [{\citenamefont {Yavuz}\ \emph {et~al.}(2016)\citenamefont {Yavuz},
  \citenamefont {Ciappina}, \citenamefont {Chac\'on}, \citenamefont {Altun},
  \citenamefont {Kling},\ and\ \citenamefont
  {Lewenstein}}]{PhysRevA.93.033404}%
  \BibitemOpen
  \bibfield  {author} {\bibinfo {author} {\bibfnamefont {I.}~\bibnamefont
  {Yavuz}}, \bibinfo {author} {\bibfnamefont {M.~F.}\ \bibnamefont {Ciappina}},
  \bibinfo {author} {\bibfnamefont {A.}~\bibnamefont {Chac\'on}}, \bibinfo
  {author} {\bibfnamefont {Z.}~\bibnamefont {Altun}}, \bibinfo {author}
  {\bibfnamefont {M.~F.}\ \bibnamefont {Kling}},\ and\ \bibinfo {author}
  {\bibfnamefont {M.}~\bibnamefont {Lewenstein}},\ }\href
  {https://doi.org/10.1103/PhysRevA.93.033404} {\bibfield  {journal} {\bibinfo
  {journal} {Phys. Rev. A}\ }\textbf {\bibinfo {volume} {93}},\ \bibinfo
  {pages} {033404} (\bibinfo {year} {2016})}\BibitemShut {NoStop}%
\bibitem [{\citenamefont {Rajpoot}\ \emph {et~al.}(2020)\citenamefont
  {Rajpoot}, \citenamefont {Holkundkar},\ and\ \citenamefont
  {Bandyopadhyay}}]{Rajpoot_2020}%
  \BibitemOpen
  \bibfield  {author} {\bibinfo {author} {\bibfnamefont {R.}~\bibnamefont
  {Rajpoot}}, \bibinfo {author} {\bibfnamefont {A.~R.}\ \bibnamefont
  {Holkundkar}},\ and\ \bibinfo {author} {\bibfnamefont {J.~N.}\ \bibnamefont
  {Bandyopadhyay}},\ }\href {https://doi.org/10.1088/1361-6455/abacd5}
  {\bibfield  {journal} {\bibinfo  {journal} {J. Phys. B: At. Mol. Opt. Phys.}\
  }\textbf {\bibinfo {volume} {53}},\ \bibinfo {pages} {205404} (\bibinfo
  {year} {2020})}\BibitemShut {NoStop}%
\bibitem [{\citenamefont {Holkundkar}\ and\ \citenamefont
  {Rajpoot}(2020)}]{Holkundkar_2020}%
  \BibitemOpen
  \bibfield  {author} {\bibinfo {author} {\bibfnamefont {A.~R.}\ \bibnamefont
  {Holkundkar}}\ and\ \bibinfo {author} {\bibfnamefont {R.}~\bibnamefont
  {Rajpoot}},\ }\href {https://doi.org/10.1088/1402-4896/aba474} {\bibfield
  {journal} {\bibinfo  {journal} {Phys. Scr.}\ }\textbf {\bibinfo {volume}
  {95}},\ \bibinfo {pages} {085607} (\bibinfo {year} {2020})}\BibitemShut
  {NoStop}%
\bibitem [{\citenamefont {Hern\'andez-Garc\'{\i}a}\ and\ \citenamefont
  {Plaja}(2016)}]{Garcia2016}%
  \BibitemOpen
  \bibfield  {author} {\bibinfo {author} {\bibfnamefont {C.}~\bibnamefont
  {Hern\'andez-Garc\'{\i}a}}\ and\ \bibinfo {author} {\bibfnamefont
  {L.}~\bibnamefont {Plaja}},\ }\href
  {https://doi.org/10.1103/PhysRevA.93.023402} {\bibfield  {journal} {\bibinfo
  {journal} {Phys. Rev. A}\ }\textbf {\bibinfo {volume} {93}},\ \bibinfo
  {pages} {023402} (\bibinfo {year} {2016})}\BibitemShut {NoStop}%
\bibitem [{\citenamefont {He}\ \emph {et~al.}(2014)\citenamefont {He},
  \citenamefont {Li}, \citenamefont {Wang}, \citenamefont {Zhang},
  \citenamefont {Lan},\ and\ \citenamefont {Lu}}]{Lixin2014}%
  \BibitemOpen
  \bibfield  {author} {\bibinfo {author} {\bibfnamefont {L.}~\bibnamefont
  {He}}, \bibinfo {author} {\bibfnamefont {Y.}~\bibnamefont {Li}}, \bibinfo
  {author} {\bibfnamefont {Z.}~\bibnamefont {Wang}}, \bibinfo {author}
  {\bibfnamefont {Q.}~\bibnamefont {Zhang}}, \bibinfo {author} {\bibfnamefont
  {P.}~\bibnamefont {Lan}},\ and\ \bibinfo {author} {\bibfnamefont
  {P.}~\bibnamefont {Lu}},\ }\href {https://doi.org/10.1103/PhysRevA.89.053417}
  {\bibfield  {journal} {\bibinfo  {journal} {Phys. Rev. A}\ }\textbf {\bibinfo
  {volume} {89}},\ \bibinfo {pages} {053417} (\bibinfo {year}
  {2014})}\BibitemShut {NoStop}%
\bibitem [{\citenamefont {Zhang}\ \emph {et~al.}(2016)\citenamefont {Zhang},
  \citenamefont {Pei}, \citenamefont {Xia}, \citenamefont {Jia},\ and\
  \citenamefont {Miao}}]{Zhang_2016}%
  \BibitemOpen
  \bibfield  {author} {\bibinfo {author} {\bibfnamefont {C.-P.}\ \bibnamefont
  {Zhang}}, \bibinfo {author} {\bibfnamefont {Y.-N.}\ \bibnamefont {Pei}},
  \bibinfo {author} {\bibfnamefont {C.-L.}\ \bibnamefont {Xia}}, \bibinfo
  {author} {\bibfnamefont {X.-F.}\ \bibnamefont {Jia}},\ and\ \bibinfo {author}
  {\bibfnamefont {X.-Y.}\ \bibnamefont {Miao}},\ }\href
  {https://doi.org/10.1088/1612-202x/14/1/015301} {\bibfield  {journal}
  {\bibinfo  {journal} {Laser Phys. Lett.}\ }\textbf {\bibinfo {volume} {14}},\
  \bibinfo {pages} {015301} (\bibinfo {year} {2016})}\BibitemShut {NoStop}%
\bibitem [{\citenamefont {Rodr\'{\i}guez-Hern\'andez}\ \emph
  {et~al.}(2022)\citenamefont {Rodr\'{\i}guez-Hern\'andez}, \citenamefont
  {Grossmann},\ and\ \citenamefont {Rost}}]{Rost2022}%
  \BibitemOpen
  \bibfield  {author} {\bibinfo {author} {\bibfnamefont {F.}~\bibnamefont
  {Rodr\'{\i}guez-Hern\'andez}}, \bibinfo {author} {\bibfnamefont
  {F.}~\bibnamefont {Grossmann}},\ and\ \bibinfo {author} {\bibfnamefont
  {J.~M.}\ \bibnamefont {Rost}},\ }\href
  {https://doi.org/10.1103/PhysRevA.105.L051102} {\bibfield  {journal}
  {\bibinfo  {journal} {Phys. Rev. A}\ }\textbf {\bibinfo {volume} {105}},\
  \bibinfo {pages} {L051102} (\bibinfo {year} {2022})}\BibitemShut {NoStop}%
\bibitem [{\citenamefont {Takahashi}\ \emph {et~al.}(2013)\citenamefont
  {Takahashi}, \citenamefont {Lan}, \citenamefont {M{\"u}cke}, \citenamefont
  {Nabekawa},\ and\ \citenamefont {Midorikawa}}]{Takahashi-2013}%
  \BibitemOpen
  \bibfield  {author} {\bibinfo {author} {\bibfnamefont {E.~J.}\ \bibnamefont
  {Takahashi}}, \bibinfo {author} {\bibfnamefont {P.}~\bibnamefont {Lan}},
  \bibinfo {author} {\bibfnamefont {O.~D.}\ \bibnamefont {M{\"u}cke}}, \bibinfo
  {author} {\bibfnamefont {Y.}~\bibnamefont {Nabekawa}},\ and\ \bibinfo
  {author} {\bibfnamefont {K.}~\bibnamefont {Midorikawa}},\ }\href
  {https://doi.org/10.1038/ncomms3691} {\bibfield  {journal} {\bibinfo
  {journal} {Nat. Commun.}\ }\textbf {\bibinfo {volume} {4}},\ \bibinfo {pages}
  {2691} (\bibinfo {year} {2013})}\BibitemShut {NoStop}%
\bibitem [{\citenamefont {Emma}\ \emph {et~al.}(2004)\citenamefont {Emma},
  \citenamefont {Bane}, \citenamefont {Cornacchia}, \citenamefont {Huang},
  \citenamefont {Schlarb}, \citenamefont {Stupakov},\ and\ \citenamefont
  {Walz}}]{Emma-2004}%
  \BibitemOpen
  \bibfield  {author} {\bibinfo {author} {\bibfnamefont {P.}~\bibnamefont
  {Emma}}, \bibinfo {author} {\bibfnamefont {K.}~\bibnamefont {Bane}}, \bibinfo
  {author} {\bibfnamefont {M.}~\bibnamefont {Cornacchia}}, \bibinfo {author}
  {\bibfnamefont {Z.}~\bibnamefont {Huang}}, \bibinfo {author} {\bibfnamefont
  {H.}~\bibnamefont {Schlarb}}, \bibinfo {author} {\bibfnamefont
  {G.}~\bibnamefont {Stupakov}},\ and\ \bibinfo {author} {\bibfnamefont
  {D.}~\bibnamefont {Walz}},\ }\href
  {https://doi.org/10.1103/PhysRevLett.92.074801} {\bibfield  {journal}
  {\bibinfo  {journal} {Phys. Rev. Lett.}\ }\textbf {\bibinfo {volume} {92}},\
  \bibinfo {pages} {074801} (\bibinfo {year} {2004})}\BibitemShut {NoStop}%
\bibitem [{\citenamefont {Ding}\ \emph {et~al.}(2015)\citenamefont {Ding},
  \citenamefont {Behrens}, \citenamefont {Coffee}, \citenamefont {Decker},
  \citenamefont {Emma}, \citenamefont {Field}, \citenamefont {Helml},
  \citenamefont {Huang}, \citenamefont {Krejcik}, \citenamefont {Krzywinski},
  \citenamefont {Loos}, \citenamefont {Lutman}, \citenamefont {Marinelli},
  \citenamefont {Maxwell},\ and\ \citenamefont {Turner}}]{Ding-2015}%
  \BibitemOpen
  \bibfield  {author} {\bibinfo {author} {\bibfnamefont {Y.}~\bibnamefont
  {Ding}}, \bibinfo {author} {\bibfnamefont {C.}~\bibnamefont {Behrens}},
  \bibinfo {author} {\bibfnamefont {R.}~\bibnamefont {Coffee}}, \bibinfo
  {author} {\bibfnamefont {F.-J.}\ \bibnamefont {Decker}}, \bibinfo {author}
  {\bibfnamefont {P.}~\bibnamefont {Emma}}, \bibinfo {author} {\bibfnamefont
  {C.}~\bibnamefont {Field}}, \bibinfo {author} {\bibfnamefont
  {W.}~\bibnamefont {Helml}}, \bibinfo {author} {\bibfnamefont
  {Z.}~\bibnamefont {Huang}}, \bibinfo {author} {\bibfnamefont
  {P.}~\bibnamefont {Krejcik}}, \bibinfo {author} {\bibfnamefont
  {J.}~\bibnamefont {Krzywinski}}, \bibinfo {author} {\bibfnamefont
  {H.}~\bibnamefont {Loos}}, \bibinfo {author} {\bibfnamefont {A.}~\bibnamefont
  {Lutman}}, \bibinfo {author} {\bibfnamefont {A.}~\bibnamefont {Marinelli}},
  \bibinfo {author} {\bibfnamefont {T.~J.}\ \bibnamefont {Maxwell}},\ and\
  \bibinfo {author} {\bibfnamefont {J.}~\bibnamefont {Turner}},\ }\href
  {https://doi.org/10.1063/1.4935429} {\bibfield  {journal} {\bibinfo
  {journal} {Appl. Phys. Lett.}\ }\textbf {\bibinfo {volume} {107}},\ \bibinfo
  {pages} {191104} (\bibinfo {year} {2015})}\BibitemShut {NoStop}%
\bibitem [{\citenamefont {Rossi}\ \emph {et~al.}(2020)\citenamefont {Rossi},
  \citenamefont {Mainz}, \citenamefont {Yang}, \citenamefont {Scheiba},
  \citenamefont {Silva-Toledo}, \citenamefont {Chia}, \citenamefont {Keathley},
  \citenamefont {Fang}, \citenamefont {M{\"u}cke}, \citenamefont {Manzoni},
  \citenamefont {Cerullo}, \citenamefont {Cirmi},\ and\ \citenamefont
  {K{\"a}rtner}}]{Rossi2020}%
  \BibitemOpen
  \bibfield  {author} {\bibinfo {author} {\bibfnamefont {G.~M.}\ \bibnamefont
  {Rossi}}, \bibinfo {author} {\bibfnamefont {R.~E.}\ \bibnamefont {Mainz}},
  \bibinfo {author} {\bibfnamefont {Y.}~\bibnamefont {Yang}}, \bibinfo {author}
  {\bibfnamefont {F.}~\bibnamefont {Scheiba}}, \bibinfo {author} {\bibfnamefont
  {M.~A.}\ \bibnamefont {Silva-Toledo}}, \bibinfo {author} {\bibfnamefont
  {S.-H.}\ \bibnamefont {Chia}}, \bibinfo {author} {\bibfnamefont {P.~D.}\
  \bibnamefont {Keathley}}, \bibinfo {author} {\bibfnamefont {S.}~\bibnamefont
  {Fang}}, \bibinfo {author} {\bibfnamefont {O.~D.}\ \bibnamefont {M{\"u}cke}},
  \bibinfo {author} {\bibfnamefont {C.}~\bibnamefont {Manzoni}}, \bibinfo
  {author} {\bibfnamefont {G.}~\bibnamefont {Cerullo}}, \bibinfo {author}
  {\bibfnamefont {G.}~\bibnamefont {Cirmi}},\ and\ \bibinfo {author}
  {\bibfnamefont {F.~X.}\ \bibnamefont {K{\"a}rtner}},\ }\href
  {https://doi.org/10.1038/s41566-020-0659-0} {\bibfield  {journal} {\bibinfo
  {journal} {Nature Photonics}\ }\textbf {\bibinfo {volume} {14}},\ \bibinfo
  {pages} {629} (\bibinfo {year} {2020})}\BibitemShut {NoStop}%
\bibitem [{\citenamefont {Chu}\ \emph {et~al.}(2016)\citenamefont {Chu},
  \citenamefont {Travers},\ and\ \citenamefont {Russell}}]{Chu_2016}%
  \BibitemOpen
  \bibfield  {author} {\bibinfo {author} {\bibfnamefont {W.-C.}\ \bibnamefont
  {Chu}}, \bibinfo {author} {\bibfnamefont {J.~C.}\ \bibnamefont {Travers}},\
  and\ \bibinfo {author} {\bibfnamefont {P.~S.~J.}\ \bibnamefont {Russell}},\
  }\href {https://doi.org/10.1088/1367-2630/18/2/023018} {\bibfield  {journal}
  {\bibinfo  {journal} {New J. Phys.}\ }\textbf {\bibinfo {volume} {18}},\
  \bibinfo {pages} {023018} (\bibinfo {year} {2016})}\BibitemShut {NoStop}%
\bibitem [{\citenamefont {Liang}\ \emph {et~al.}(2017)\citenamefont {Liang},
  \citenamefont {Krogen}, \citenamefont {Wang}, \citenamefont {Park},
  \citenamefont {Kroh}, \citenamefont {Zawilski}, \citenamefont {Schunemann},
  \citenamefont {Moses}, \citenamefont {DiMauro}, \citenamefont {K{\"a}rtner},\
  and\ \citenamefont {Hong}}]{Liang2017}%
  \BibitemOpen
  \bibfield  {author} {\bibinfo {author} {\bibfnamefont {H.}~\bibnamefont
  {Liang}}, \bibinfo {author} {\bibfnamefont {P.}~\bibnamefont {Krogen}},
  \bibinfo {author} {\bibfnamefont {Z.}~\bibnamefont {Wang}}, \bibinfo {author}
  {\bibfnamefont {H.}~\bibnamefont {Park}}, \bibinfo {author} {\bibfnamefont
  {T.}~\bibnamefont {Kroh}}, \bibinfo {author} {\bibfnamefont {K.}~\bibnamefont
  {Zawilski}}, \bibinfo {author} {\bibfnamefont {P.}~\bibnamefont
  {Schunemann}}, \bibinfo {author} {\bibfnamefont {J.}~\bibnamefont {Moses}},
  \bibinfo {author} {\bibfnamefont {L.~F.}\ \bibnamefont {DiMauro}}, \bibinfo
  {author} {\bibfnamefont {F.~X.}\ \bibnamefont {K{\"a}rtner}},\ and\ \bibinfo
  {author} {\bibfnamefont {K.-H.}\ \bibnamefont {Hong}},\ }\href
  {https://doi.org/10.1038/s41467-017-00193-4} {\bibfield  {journal} {\bibinfo
  {journal} {Nat. Commun.}\ }\textbf {\bibinfo {volume} {8}},\ \bibinfo {pages}
  {141} (\bibinfo {year} {2017})}\BibitemShut {NoStop}%
\bibitem [{\citenamefont {Tate}\ \emph {et~al.}(2007)\citenamefont {Tate},
  \citenamefont {Auguste}, \citenamefont {Muller}, \citenamefont {Sali\`eres},
  \citenamefont {Agostini},\ and\ \citenamefont {DiMauro}}]{Tate2007_scale}%
  \BibitemOpen
  \bibfield  {author} {\bibinfo {author} {\bibfnamefont {J.}~\bibnamefont
  {Tate}}, \bibinfo {author} {\bibfnamefont {T.}~\bibnamefont {Auguste}},
  \bibinfo {author} {\bibfnamefont {H.~G.}\ \bibnamefont {Muller}}, \bibinfo
  {author} {\bibfnamefont {P.}~\bibnamefont {Sali\`eres}}, \bibinfo {author}
  {\bibfnamefont {P.}~\bibnamefont {Agostini}},\ and\ \bibinfo {author}
  {\bibfnamefont {L.~F.}\ \bibnamefont {DiMauro}},\ }\href
  {https://doi.org/10.1103/PhysRevLett.98.013901} {\bibfield  {journal}
  {\bibinfo  {journal} {Phys. Rev. Lett.}\ }\textbf {\bibinfo {volume} {98}},\
  \bibinfo {pages} {013901} (\bibinfo {year} {2007})}\BibitemShut {NoStop}%
\bibitem [{\citenamefont {Wang}\ \emph {et~al.}(2014)\citenamefont {Wang},
  \citenamefont {Yu}, \citenamefont {Shi}, \citenamefont {Zhang}, \citenamefont
  {Cao}, \citenamefont {Jiang},\ and\ \citenamefont {Lu}}]{wang2014_scale}%
  \BibitemOpen
  \bibfield  {author} {\bibinfo {author} {\bibfnamefont {Y.}~\bibnamefont
  {Wang}}, \bibinfo {author} {\bibfnamefont {C.}~\bibnamefont {Yu}}, \bibinfo
  {author} {\bibfnamefont {Q.}~\bibnamefont {Shi}}, \bibinfo {author}
  {\bibfnamefont {Y.}~\bibnamefont {Zhang}}, \bibinfo {author} {\bibfnamefont
  {X.}~\bibnamefont {Cao}}, \bibinfo {author} {\bibfnamefont {S.}~\bibnamefont
  {Jiang}},\ and\ \bibinfo {author} {\bibfnamefont {R.}~\bibnamefont {Lu}},\
  }\href {https://doi.org/10.1103/PhysRevA.89.023825} {\bibfield  {journal}
  {\bibinfo  {journal} {Phys. Rev. A}\ }\textbf {\bibinfo {volume} {89}},\
  \bibinfo {pages} {023825} (\bibinfo {year} {2014})}\BibitemShut {NoStop}%
\bibitem [{\citenamefont {Emelina}\ \emph {et~al.}(2019)\citenamefont
  {Emelina}, \citenamefont {Emelin},\ and\ \citenamefont
  {Ryabikin}}]{Emelina2019}%
  \BibitemOpen
  \bibfield  {author} {\bibinfo {author} {\bibfnamefont {A.~S.}\ \bibnamefont
  {Emelina}}, \bibinfo {author} {\bibfnamefont {M.~Y.}\ \bibnamefont
  {Emelin}},\ and\ \bibinfo {author} {\bibfnamefont {M.~Y.}\ \bibnamefont
  {Ryabikin}},\ }\href {https://doi.org/10.1364/JOSAB.36.003236} {\bibfield
  {journal} {\bibinfo  {journal} {J. Opt. Soc. Am. B}\ }\textbf {\bibinfo
  {volume} {36}},\ \bibinfo {pages} {3236} (\bibinfo {year}
  {2019})}\BibitemShut {NoStop}%
\bibitem [{\citenamefont {Schiessl}\ \emph {et~al.}(2007)\citenamefont
  {Schiessl}, \citenamefont {Ishikawa}, \citenamefont {Persson},\ and\
  \citenamefont {Burgd\"orfer}}]{Schiessl2007_scale}%
  \BibitemOpen
  \bibfield  {author} {\bibinfo {author} {\bibfnamefont {K.}~\bibnamefont
  {Schiessl}}, \bibinfo {author} {\bibfnamefont {K.~L.}\ \bibnamefont
  {Ishikawa}}, \bibinfo {author} {\bibfnamefont {E.}~\bibnamefont {Persson}},\
  and\ \bibinfo {author} {\bibfnamefont {J.}~\bibnamefont {Burgd\"orfer}},\
  }\href {https://doi.org/10.1103/PhysRevLett.99.253903} {\bibfield  {journal}
  {\bibinfo  {journal} {Phys. Rev. Lett.}\ }\textbf {\bibinfo {volume} {99}},\
  \bibinfo {pages} {253903} (\bibinfo {year} {2007})}\BibitemShut {NoStop}%
\bibitem [{\citenamefont {Ishikawa}\ \emph {et~al.}(2009)\citenamefont
  {Ishikawa}, \citenamefont {Schiessl}, \citenamefont {Persson},\ and\
  \citenamefont {Burgd\"orfer}}]{Ishikawa2009_scale}%
  \BibitemOpen
  \bibfield  {author} {\bibinfo {author} {\bibfnamefont {K.~L.}\ \bibnamefont
  {Ishikawa}}, \bibinfo {author} {\bibfnamefont {K.}~\bibnamefont {Schiessl}},
  \bibinfo {author} {\bibfnamefont {E.}~\bibnamefont {Persson}},\ and\ \bibinfo
  {author} {\bibfnamefont {J.}~\bibnamefont {Burgd\"orfer}},\ }\href
  {https://doi.org/10.1103/PhysRevA.79.033411} {\bibfield  {journal} {\bibinfo
  {journal} {Phys. Rev. A}\ }\textbf {\bibinfo {volume} {79}},\ \bibinfo
  {pages} {033411} (\bibinfo {year} {2009})}\BibitemShut {NoStop}%
\bibitem [{\citenamefont {Nefedova}\ \emph {et~al.}(2018)\citenamefont
  {Nefedova}, \citenamefont {Ciappina}, \citenamefont {Finke}, \citenamefont
  {Albrecht}, \citenamefont {Kozlová},\ and\ \citenamefont
  {Nejdl}}]{Nefedova2018}%
  \BibitemOpen
  \bibfield  {author} {\bibinfo {author} {\bibfnamefont {V.~E.}\ \bibnamefont
  {Nefedova}}, \bibinfo {author} {\bibfnamefont {M.~F.}\ \bibnamefont
  {Ciappina}}, \bibinfo {author} {\bibfnamefont {O.}~\bibnamefont {Finke}},
  \bibinfo {author} {\bibfnamefont {M.}~\bibnamefont {Albrecht}}, \bibinfo
  {author} {\bibfnamefont {M.}~\bibnamefont {Kozlová}},\ and\ \bibinfo
  {author} {\bibfnamefont {J.}~\bibnamefont {Nejdl}},\ }\href
  {https://doi.org/10.1063/1.5050691} {\bibfield  {journal} {\bibinfo
  {journal} {Applied Physics Letters}\ }\textbf {\bibinfo {volume} {113}},\
  \bibinfo {pages} {191101} (\bibinfo {year} {2018})}\BibitemShut {NoStop}%
\bibitem [{\citenamefont {Frolov}\ \emph {et~al.}(2009)\citenamefont {Frolov},
  \citenamefont {Manakov}, \citenamefont {Sarantseva}, \citenamefont {Emelin},
  \citenamefont {Ryabikin},\ and\ \citenamefont {Starace}}]{Frolov2009_scale}%
  \BibitemOpen
  \bibfield  {author} {\bibinfo {author} {\bibfnamefont {M.~V.}\ \bibnamefont
  {Frolov}}, \bibinfo {author} {\bibfnamefont {N.~L.}\ \bibnamefont {Manakov}},
  \bibinfo {author} {\bibfnamefont {T.~S.}\ \bibnamefont {Sarantseva}},
  \bibinfo {author} {\bibfnamefont {M.~Y.}\ \bibnamefont {Emelin}}, \bibinfo
  {author} {\bibfnamefont {M.~Y.}\ \bibnamefont {Ryabikin}},\ and\ \bibinfo
  {author} {\bibfnamefont {A.~F.}\ \bibnamefont {Starace}},\ }\href
  {https://doi.org/10.1103/PhysRevLett.102.243901} {\bibfield  {journal}
  {\bibinfo  {journal} {Phys. Rev. Lett.}\ }\textbf {\bibinfo {volume} {102}},\
  \bibinfo {pages} {243901} (\bibinfo {year} {2009})}\BibitemShut {NoStop}%
\bibitem [{\citenamefont {Marceau}\ \emph {et~al.}(2017)\citenamefont
  {Marceau}, \citenamefont {Hammond}, \citenamefont {Naumov}, \citenamefont
  {Corkum},\ and\ \citenamefont {Villeneuve}}]{Marceau_2017}%
  \BibitemOpen
  \bibfield  {author} {\bibinfo {author} {\bibfnamefont {C.}~\bibnamefont
  {Marceau}}, \bibinfo {author} {\bibfnamefont {T.~J.}\ \bibnamefont
  {Hammond}}, \bibinfo {author} {\bibfnamefont {A.~Y.}\ \bibnamefont {Naumov}},
  \bibinfo {author} {\bibfnamefont {P.~B.}\ \bibnamefont {Corkum}},\ and\
  \bibinfo {author} {\bibfnamefont {D.~M.}\ \bibnamefont {Villeneuve}},\ }\href
  {https://doi.org/10.1088/2399-6528/aa74f6} {\bibfield  {journal} {\bibinfo
  {journal} {J. Phys. Commun.}\ }\textbf {\bibinfo {volume} {1}},\ \bibinfo
  {pages} {015009} (\bibinfo {year} {2017})}\BibitemShut {NoStop}%
\bibitem [{\citenamefont {Lin}\ \emph {et~al.}(2006)\citenamefont {Lin},
  \citenamefont {Zheng},\ and\ \citenamefont {Becker}}]{LinSubcycle-2006}%
  \BibitemOpen
  \bibfield  {author} {\bibinfo {author} {\bibfnamefont {Q.}~\bibnamefont
  {Lin}}, \bibinfo {author} {\bibfnamefont {J.}~\bibnamefont {Zheng}},\ and\
  \bibinfo {author} {\bibfnamefont {W.}~\bibnamefont {Becker}},\ }\href
  {https://doi.org/10.1103/PhysRevLett.97.253902} {\bibfield  {journal}
  {\bibinfo  {journal} {Phys. Rev. Lett.}\ }\textbf {\bibinfo {volume} {97}},\
  \bibinfo {pages} {253902} (\bibinfo {year} {2006})}\BibitemShut {NoStop}%
\bibitem [{\citenamefont {Zheng}\ \emph {et~al.}(2011)\citenamefont {Zheng},
  \citenamefont {Qiu},\ and\ \citenamefont {Lin}}]{Zheng_2011}%
  \BibitemOpen
  \bibfield  {author} {\bibinfo {author} {\bibfnamefont {J.}~\bibnamefont
  {Zheng}}, \bibinfo {author} {\bibfnamefont {E.}~\bibnamefont {Qiu}},\ and\
  \bibinfo {author} {\bibfnamefont {Q.}~\bibnamefont {Lin}},\ }\href
  {https://doi.org/10.1088/2040-8978/13/7/075206} {\bibfield  {journal}
  {\bibinfo  {journal} {J. Opt.}\ }\textbf {\bibinfo {volume} {13}},\ \bibinfo
  {pages} {075206} (\bibinfo {year} {2011})}\BibitemShut {NoStop}%
\bibitem [{\citenamefont {Tong}\ and\ \citenamefont {Chu}(1997)}]{TONG1997119}%
  \BibitemOpen
  \bibfield  {author} {\bibinfo {author} {\bibfnamefont {X.-M.}\ \bibnamefont
  {Tong}}\ and\ \bibinfo {author} {\bibfnamefont {S.-I.}\ \bibnamefont {Chu}},\
  }\href {https://doi.org/https://doi.org/10.1016/S0301-0104(97)00063-3}
  {\bibfield  {journal} {\bibinfo  {journal} {Chem. Phys.}\ }\textbf {\bibinfo
  {volume} {217}},\ \bibinfo {pages} {119} (\bibinfo {year}
  {1997})}\BibitemShut {NoStop}%
\bibitem [{\citenamefont {Wang}\ \emph {et~al.}(2003)\citenamefont {Wang},
  \citenamefont {Lin},\ and\ \citenamefont {Wang}}]{PhysRevE.67.016503}%
  \BibitemOpen
  \bibfield  {author} {\bibinfo {author} {\bibfnamefont {Z.}~\bibnamefont
  {Wang}}, \bibinfo {author} {\bibfnamefont {Q.}~\bibnamefont {Lin}},\ and\
  \bibinfo {author} {\bibfnamefont {Z.}~\bibnamefont {Wang}},\ }\href
  {https://doi.org/10.1103/PhysRevE.67.016503} {\bibfield  {journal} {\bibinfo
  {journal} {Phys. Rev. E}\ }\textbf {\bibinfo {volume} {67}},\ \bibinfo
  {pages} {016503} (\bibinfo {year} {2003})}\BibitemShut {NoStop}%
\bibitem [{\citenamefont {Heyman}\ and\ \citenamefont
  {Felsen}(1989)}]{Heyman-89}%
  \BibitemOpen
  \bibfield  {author} {\bibinfo {author} {\bibfnamefont {E.}~\bibnamefont
  {Heyman}}\ and\ \bibinfo {author} {\bibfnamefont {L.~B.}\ \bibnamefont
  {Felsen}},\ }\href {https://doi.org/10.1364/JOSAA.6.000806} {\bibfield
  {journal} {\bibinfo  {journal} {J. Opt. Soc. Am. A}\ }\textbf {\bibinfo
  {volume} {6}},\ \bibinfo {pages} {806} (\bibinfo {year} {1989})}\BibitemShut
  {NoStop}%
\bibitem [{\citenamefont {Tong}\ and\ \citenamefont {Lin}(2005)}]{Tong_2005}%
  \BibitemOpen
  \bibfield  {author} {\bibinfo {author} {\bibfnamefont {X.~M.}\ \bibnamefont
  {Tong}}\ and\ \bibinfo {author} {\bibfnamefont {C.~D.}\ \bibnamefont {Lin}},\
  }\href {https://doi.org/10.1088/0953-4075/38/15/001} {\bibfield  {journal}
  {\bibinfo  {journal} {J. Phys. B: At. Mol. Opt. Phys.}\ }\textbf {\bibinfo
  {volume} {38}},\ \bibinfo {pages} {2593} (\bibinfo {year}
  {2005})}\BibitemShut {NoStop}%
\bibitem [{\citenamefont {van~de Sand}\ and\ \citenamefont
  {Rost}(1999)}]{sandPRL_1999}%
  \BibitemOpen
  \bibfield  {author} {\bibinfo {author} {\bibfnamefont {G.}~\bibnamefont
  {van~de Sand}}\ and\ \bibinfo {author} {\bibfnamefont {J.~M.}\ \bibnamefont
  {Rost}},\ }\href {https://doi.org/10.1103/PhysRevLett.83.524} {\bibfield
  {journal} {\bibinfo  {journal} {Phys. Rev. Lett.}\ }\textbf {\bibinfo
  {volume} {83}},\ \bibinfo {pages} {524} (\bibinfo {year} {1999})}\BibitemShut
  {NoStop}%
\bibitem [{\citenamefont {Peng}\ \emph {et~al.}(2020)\citenamefont {Peng},
  \citenamefont {Starace}, \citenamefont {Shao},\ and\ \citenamefont
  {Djiokap}}]{Peng-2020}%
  \BibitemOpen
  \bibfield  {author} {\bibinfo {author} {\bibfnamefont {D.}~\bibnamefont
  {Peng}}, \bibinfo {author} {\bibfnamefont {A.~F.}\ \bibnamefont {Starace}},
  \bibinfo {author} {\bibfnamefont {H.-C.}\ \bibnamefont {Shao}},\ and\
  \bibinfo {author} {\bibfnamefont {J.~M.~N.}\ \bibnamefont {Djiokap}},\ }\href
  {https://doi.org/10.1103/PhysRevA.102.063126} {\bibfield  {journal} {\bibinfo
   {journal} {Phys. Rev. A}\ }\textbf {\bibinfo {volume} {102}},\ \bibinfo
  {pages} {063126} (\bibinfo {year} {2020})}\BibitemShut {NoStop}%
\bibitem [{\citenamefont {Sanderson}\ and\ \citenamefont
  {Curtin}(2016)}]{Sanderson2016}%
  \BibitemOpen
  \bibfield  {author} {\bibinfo {author} {\bibfnamefont {C.}~\bibnamefont
  {Sanderson}}\ and\ \bibinfo {author} {\bibfnamefont {R.}~\bibnamefont
  {Curtin}},\ }\href {https://doi.org/10.21105/joss.00026} {\bibfield
  {journal} {\bibinfo  {journal} {Journal of Open Source Software}\ }\textbf
  {\bibinfo {volume} {1}},\ \bibinfo {pages} {26} (\bibinfo {year}
  {2016})}\BibitemShut {NoStop}%
\bibitem [{\citenamefont {Frolov}\ \emph {et~al.}(2015)\citenamefont {Frolov},
  \citenamefont {Manakov}, \citenamefont {Xiong}, \citenamefont {Peng},
  \citenamefont {Burgd\"orfer},\ and\ \citenamefont {Starace}}]{Frolov2015}%
  \BibitemOpen
  \bibfield  {author} {\bibinfo {author} {\bibfnamefont {M.~V.}\ \bibnamefont
  {Frolov}}, \bibinfo {author} {\bibfnamefont {N.~L.}\ \bibnamefont {Manakov}},
  \bibinfo {author} {\bibfnamefont {W.-H.}\ \bibnamefont {Xiong}}, \bibinfo
  {author} {\bibfnamefont {L.-Y.}\ \bibnamefont {Peng}}, \bibinfo {author}
  {\bibfnamefont {J.}~\bibnamefont {Burgd\"orfer}},\ and\ \bibinfo {author}
  {\bibfnamefont {A.~F.}\ \bibnamefont {Starace}},\ }\href
  {https://doi.org/10.1103/PhysRevA.92.023409} {\bibfield  {journal} {\bibinfo
  {journal} {Phys. Rev. A}\ }\textbf {\bibinfo {volume} {92}},\ \bibinfo
  {pages} {023409} (\bibinfo {year} {2015})}\BibitemShut {NoStop}%
\bibitem [{\citenamefont {Jin}\ \emph {et~al.}(2011)\citenamefont {Jin},
  \citenamefont {Le},\ and\ \citenamefont {Lin}}]{Jin2011}%
  \BibitemOpen
  \bibfield  {author} {\bibinfo {author} {\bibfnamefont {C.}~\bibnamefont
  {Jin}}, \bibinfo {author} {\bibfnamefont {A.-T.}\ \bibnamefont {Le}},\ and\
  \bibinfo {author} {\bibfnamefont {C.~D.}\ \bibnamefont {Lin}},\ }\href
  {https://doi.org/10.1103/PhysRevA.83.023411} {\bibfield  {journal} {\bibinfo
  {journal} {Phys. Rev. A}\ }\textbf {\bibinfo {volume} {83}},\ \bibinfo
  {pages} {023411} (\bibinfo {year} {2011})}\BibitemShut {NoStop}%
\end{thebibliography}

%

\end{document}